\documentclass[a4paper]{article}

\usepackage[T1]{fontenc} 
\usepackage{float}       
\usepackage{helvet}      
\usepackage{mathptmx}    
\usepackage{rsc}         
\usepackage{setspace}    
\AtBeginDocument{\doublespacing}

\newfloat{chart}{htbp}{loc}
\newfloat{graph}{htbp}{loh}
\newfloat{scheme}{htbp}{los}

\usepackage[version=3]{mhchem} 

\usepackage{graphicx}
\newcommand{\op}[1]{%
    \fontdimen12\textfont3=2pt\fontdimen12\scriptfont3=1.4pt%
    \!\null\mathop{\vphantom{#1}\smash{#1}}\limits_{\sim}\null\!}

\newcommand{\fmref}[1]{(\protect\ref{#1})}

\def\ket#1{\, | \, {#1} \, \rangle}

\author{%
  J{\"u}rgen Schnack\thanks{%
    Universit{\"a}t Bielefeld, Fakult{\"a}t f{\"u}r Physik, Postfach 100131, D-33501 Bielefeld, Germany%
  }%
}

\title{Effects of frustration on magnetic molecules:\\ a survey
  from Olivier Kahn till today}

\begin{document}

\maketitle

\begin{abstract}
  In magnetism, of which molecular magnetism is a part, the term
  frustration is used rather sloppily. Sometimes one gains the
  impression that if the reason for some phenomenon is not quite
  clear then it is attributed to frustration.  In this paper I
  discuss the effects of frustration that I feel are relevant
  for the field of molecular magnetism. As will become clear
  later they indeed lead to a variety of unusual magnetic
  properties.
\end{abstract}

\section{Introduction}

\begin{figure}[ht!]
\centering
\includegraphics[clip,width=45mm]{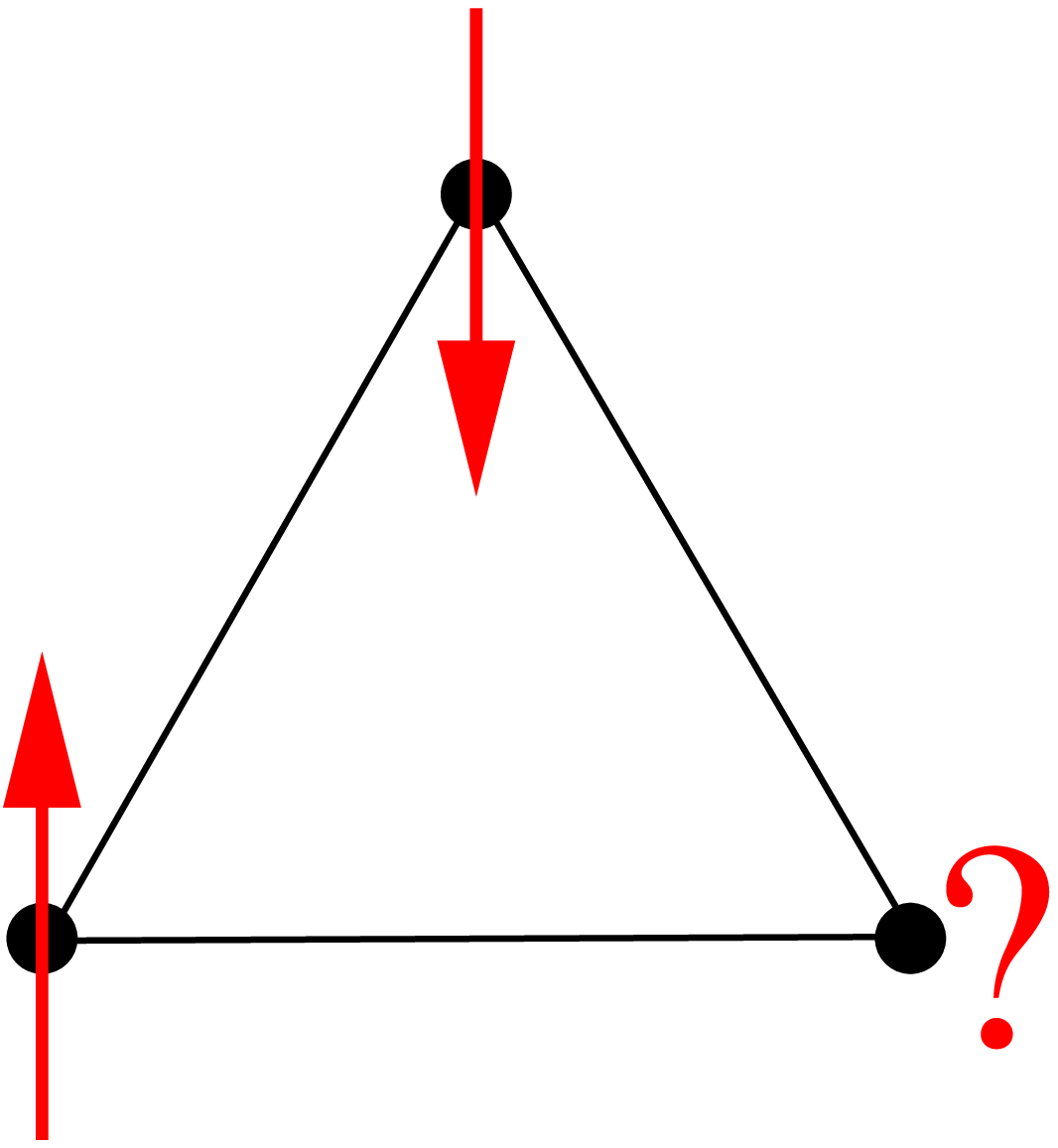}
\quad
\includegraphics[clip,width=65mm]{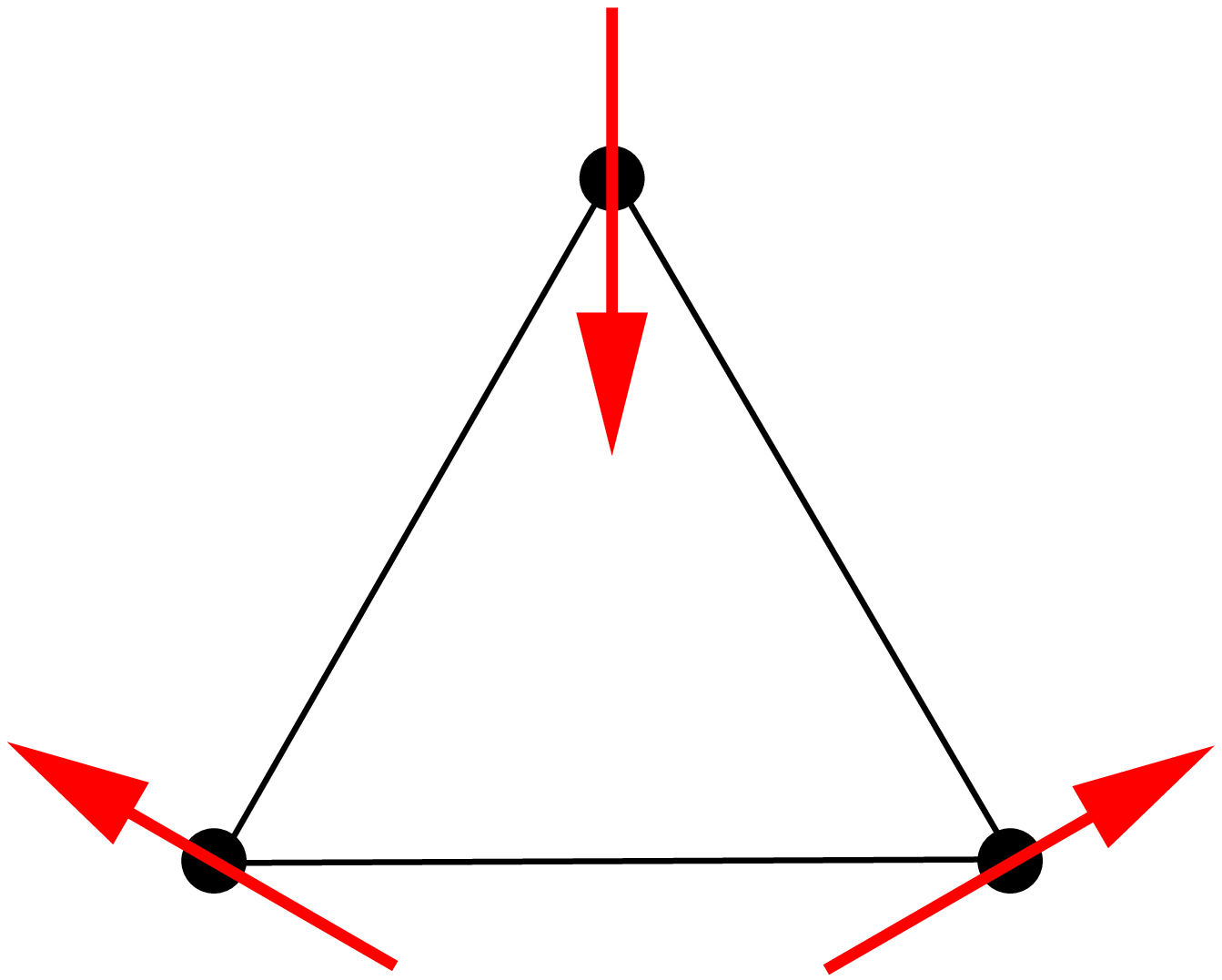}
\caption{L.h.s.: Typical sketch of a frustrated triangular
  classical spin system, where the third spin ``does not know
  where to point". R.h.s.: Classical ground state.}
\label{dt-schnack-F-1}
\end{figure}

In an article specifically devoted to the effect of frustration
on magnetic molecules Olivier Kahn demands a more thorough
discussion of frustration effects. He summarizes the
intellectual deficiencies when discussing for instance a
triangle of spins $s=1$ as follows:\cite{Kahn:CPL97} ``What
might be frustrating for some researchers is not to be able to
represent the $^1A_1$ ground state with up and down arrows."
Being a physicist I silently identify ``researcher" as
``physicist" which provides good motivation for a survey of this
phenomenon on which both chemistry and physics had their
different focus.

Following the bibliographic path backwards in history the term
frustration was introduced by P.~W.~Anderson in a private
communication and employed by G.~Toulouse and S.~Kirkpatrick in
connection with spin glasses\cite{Tou:CP77,Kir:PRB77}. In
condensed matter physics it was later transfered to the
discussion of special spin lattices as for instance the
kagom\'{e} or the pyrochlore lattice
antiferromagnets\cite{Ram:ARMS94,Gre:JMC01,Moe:CJP01,BrG:Science01}.
There are two approaches to frustration. The first one starts by
considering the graph of interactions between the participating
spins\cite{Tou:CP77}. This approach defines a system with
``competing interactions" as frustrated. The second approach
focuses on the phenomena ascribed to frustration. It is in this
context that Olivier Kahn criticizes the use of the term
frustration for systems that have competing interactions but no
resulting specific properties. To call a spin triangle of
integer spins frustrated is thus meaningless to him since it
does not show the specific property of ground state
degeneracy\cite{Kahn:CPL97}.

In this article I undertake the attempt to reconcile the various
viewpoints. I will therefore concentrate on geometric
frustration and argue that one could consider frustration as the
opposite of bipartiteness. This approach has two clear
advantages: one has a strict definition of bipartiteness and one
can discuss the frustration related properties in contrast to
the properties of bipartite systems. The article is thus
organized as follows. In section \ref{schnack-bipartite} I will
repeat the concept of bipartiteness and discuss the resulting
properties. Then the observable signatures of frustration will
be introduced in section \ref{schnack-frustration} as there
are for instance ground-state degeneracy, low-lying singlets,
non-collinear ground states, magnetization plateaus,
magnetization jumps, as well as special magnetocaloric
properties.

\section{Bipartiteness}
\label{schnack-bipartite}

Before discussing the concept of bipartiteness and the resulting
properties I define the Heisenberg model as the physical
background model for interacting spins. In the Hamiltonian
\begin{eqnarray}
\label{E-2-1}
\op{H}
&=&
-
\sum_{i,j}\;
{J}_{ij}
\op{\vec{s}}_i \cdot \op{\vec{s}}_j
+
g\, \mu_B\, B\,
\sum_{i}\;
\op{s}^z_i
\ ,
\end{eqnarray}
the first term (Heisenberg Hamiltonian) models the isotropic
exchange interaction between spins centered at sites $i$ and
$j$. ${J}_{ij}$ is the exchange parameter; a negative ${J}_{ij}$
corresponds to an antiferromagnetic coupling of the spins. The
second term (Zeeman term) represents the interaction with the
external magnetic field. For the sake of simplicity it is
assumed that all spins have the same spin quantum number
$s_1=s_2=\dots=s_N=s$ as well as the same $g$-factor.  With
these simplifications the model is SU(2) invariant, i.e. the
total spin commutes with the Hamiltonian in the following way
\begin{eqnarray}
\label{E-2-2}
\left[\op{H}, \op{\vec{S}}^2\right] = 0
\quad &\text{\&}&\quad
\left[\op{H}, \op{{S}}^z\right] = 0
\ .
\end{eqnarray}
This means that a basis exists where the basis states $\ket{\nu}$
are simultaneous eigenstates of $\op{H}, \op{\vec{S}}^2$, and
$\op{{S}}^z$, i.e. $\op{H}\ket{\nu}=E_\nu\ket{\nu},
\op{\vec{S}}^2\ket{\nu}=S_\nu(S_\nu+1)\ket{\nu}, \text{and}\
\op{{S}}_z\ket{\nu}=M_\nu\ket{\nu}$.

The concept of bipartiteness has the clear advantage that it can
be rigorously defined and that a number of consequences can be
proven. In addition, it also provides a classical meaning. In
the classical ground state of a bipartite system spins that
interact ferromagnetically are mutually aligned parallel and
spins that interact antiferromagnetically are mutually aligned
anti-parallel. Following the pioneering work of Lieb, Schultz,
and Mattis \cite{LSM:AP61,LiM:JMP62} this can now be defined
without recourse to classical spin systems.

\begin{figure}[ht!]
\centering
\includegraphics[clip,width=45mm]{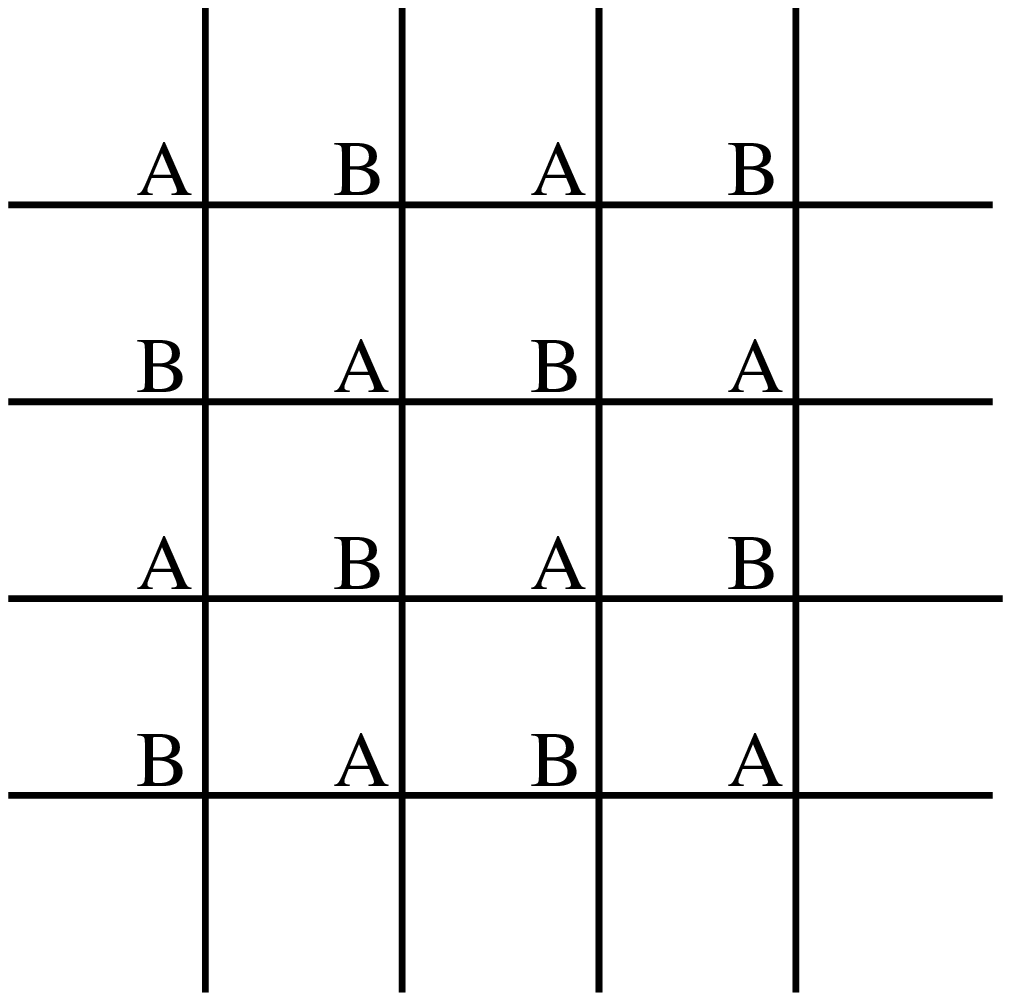}
\quad
\includegraphics[clip,width=45mm]{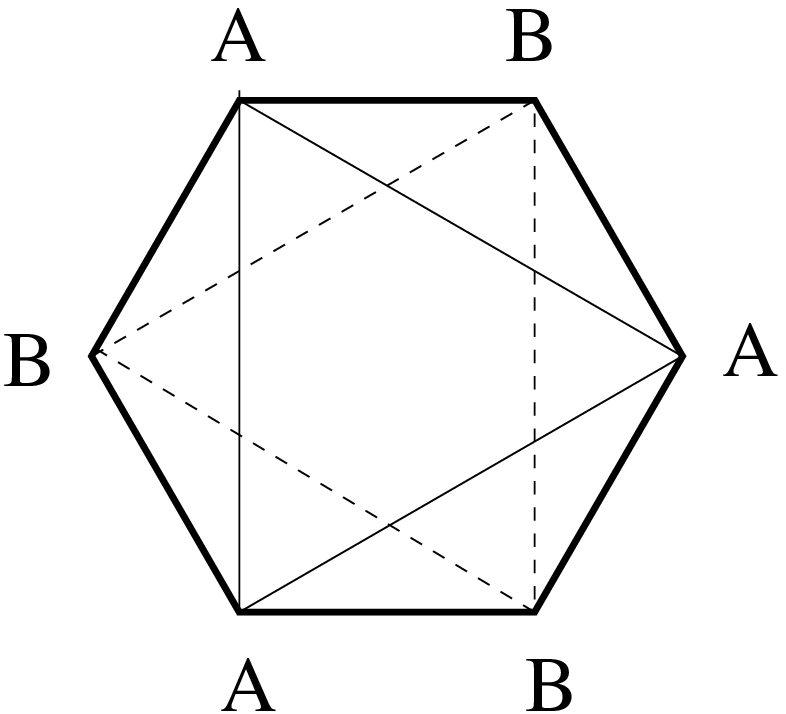}
\caption{Schematic picture of fictitious spin systems: the
  square lattice on the l.h.s. is bipartite, the hexagon on the
  r.h.s. only if the intra-$A$ and intra-$B$ sublattice interactions
  fulfill the conditions put up by Lieb, Schultz, and Mattis.}
\label{dt-schnack-F-2}
\end{figure}

If the spin system can be decomposed into subsystems
(sublattices) $A$ and $B$ such that all exchange parameters
(also those that are zero) fulfill $J(x_A, y_B) \le g^2\ ,
J(x_A, y_A) \ge g^2\ , J(x_B, y_B) \ge g^2$, the system is
called bipartite. Here $x_A$ and $y_A$ denote the sites of spins
belonging to sublattice $A$ and $x_B$ and $y_B$ the sites of
spins belonging to sublattice $B$; $g$ is a real number that has
to be found in order to establish the partition. It is essential
for this definition that antiferromagnetic interactions are
represented by negative exchange parameters.

Figure \ref{dt-schnack-F-2} shows two examples. The square
lattice on the l.h.s. is bipartite for antiferromagnetic nearest
neighbor interactions. The hexagon on the r.h.s. shows three
types of interactions. Let's assume that the interaction along
the hexagon edges is antiferromagnetic as in many spin
rings. Then the system would be bipartite if the other two
interactions that act inside the $A$ and the $B$ sublattice,
respectively, are ferromagnetic. In this case $g=0$ would allow
the given partition. If on the contrary any of the shown
interactions inside the hexagon would also be antiferromagnetic
then a valid partition cannot be found.

The proven consequences of bipartiteness are:
\begin{itemize}
\item Let ${\mathcal S}=|S_A-S_B|$, where $S_A$ and $S_B$ are the
  maximum possible spins on the sublattices $A$ and $B$. Then
  the ground state of the Heisenberg Hamiltonian belongs at most
  to total spin ${\mathcal S}$. This immediately implies that
  for $S_A=S_B$ the ground state spin is zero\cite{LiM:JMP62},
  but also explains ferrimagnetic cases.
\item For all total spins $S$ with $S\ge {\mathcal S}$ the
  minimal energies $E(S)$ in the sectors of total spin fulfill
  $E(S+1)>E(S)$. Again, for $S_A=S_B$ this holds for all total
  spin quantum numbers\cite{LiM:JMP62}. The levels of minimal
  energy are non-degenerate except for the trivial degeneracy
  related to the magnetic quantum number $M$.
\item Bipartiteness also enables one to determine phase
  relations of ground state wave functions. The most prominent
  example is the sign rule of Marshall and Peierls, that can be
  connected to the momentum quantum number of periodic systems
  such as spin rings\cite{Mar:PRS55}.
\end{itemize}

\begin{figure}[ht!]
\centering
\includegraphics[clip,width=75mm]{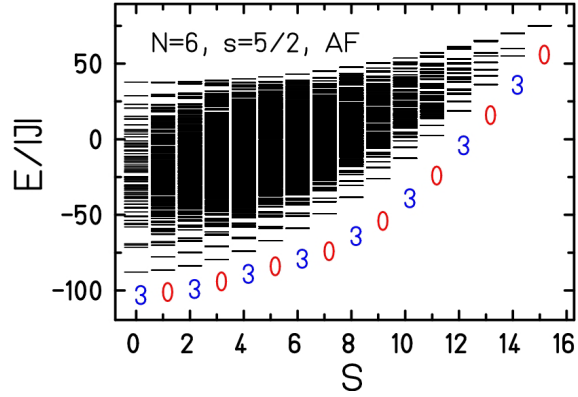}
\caption{Energy eigenvalues of a spin ring ($N=6, s=5/2$) with
  antiferromagnetic nearest neighbor coupling as function of
  total spin $S$.}
\label{dt-schnack-F-3}
\end{figure}

I would like to discuss these properties with the help of an
example. Figure~\ref{dt-schnack-F-3} shows the energy spectrum of a spin
ring ($N=6, s=5/2$) with antiferromagnetic nearest neighbor
coupling as function of total spin $S$. This system is
bipartite. One easily recognizes that the ground state belongs
to $S=0$ which it must since both sublattices possess the same
maximum possible spin of $3\times 5/2$. One also sees that 
$E(S+1)>E(S)$ strictly holds for all $S$. The system is $C_6$
symmetric, or in the language of condensed matter physics
translational invariant with periodic boundary conditions after
six sites. This gives rise to 6 irreducible representations with
complex characters, that can be expressed as 
\begin{eqnarray}
\label{E-2-3}
\exp\left\{-i \frac{2\pi k}{N} \right\}\quad , \quad
k=0,1,\dots, N
\ .
\end{eqnarray}
 $k=0,1,2,3,4,5$ labels the characters or momentum quantum
numbers, respectively. The sign rule of Marshall and Peierls
allows us to determine the momentum quantum numbers $k$ (wave
numbers) for the relative ground state in each subspace of total
spin $S$. These numbers are given in the figure close to the
lowest levels.

Besides the proven properties there are softer ones that seem to
hold for bipartite systems.
\begin{itemize}
\item The lowest levels of regular, i.e. highly symmetric,
  bipartite antiferromagnets as a function of $S$ are arranged
  in an approximate parabola (Lande interval rule, rotational
  bands)\cite{And:PR52,JJL:PRL99,ScL:PRB00,ACC:ICA00,Wal:PRB01},
  compare Fig.~\ref{dt-schnack-F-3}.
\item The lowest band can be understood as originating from an
  effective Hamiltonian where the two sublattice spins
  $S_A=Ns/2$ and $S_B=Ns/2$ interact
  antiferromagnetically\cite{And:PR52}. This picture also
  motivates the existence of a second band in which an $S=0$
  level does not exist, since $S_A=Ns/2$ and $S_B=Ns/2-1$ cannot
  be coupled to $S=0$. This effect is clearly visible in the
  spectrum shown in Fig.~\ref{dt-schnack-F-3}.
\item Since the curvature of the minimal energy function $E(S)$
  determines the magnetization curve at $T=0$ or equivalently
  the order of successive level crossings, an approximate
  quadratic dependence yields with increasing field $B$
  successive crossings of levels belonging to adjacent total
  spin quantum numbers with practically equidistant field
  spacings. This means that ($S=1, M=-1$) crosses ($S=0, M=0$)
  at $B_1$, then ($S=2, M=-2$) crosses ($S=1, M=-1$) at
  $B_2\approx 2 B_1$ and so on. The resulting low-temperature
  magnetization curve will thus be a rather regular staircase.
\end{itemize}
We will see that these properties are entirely different for
frustrated spin systems.

\section{Effects of frustration}
\label{schnack-frustration}

In this section I discuss the various effects frustration can
have on the energy spectrum and on magnetic observables. The
investigation of plain triangles or tetrahedra is not really
enlightening since both have a very simple energy spectrum
consisting of one band of energy levels. The reason is that for
both systems the Heisenberg Hamiltonian can be simplified by
completing the square, i.e. the Heisenberg term is simply
proportional to $\op{\vec{S}}^2$. Nevertheless, the degeneracy
of energy eigenvalues remains an issue\cite{Kahn:CPL97,dai:672}.

\begin{figure}[ht!]
\centering
\includegraphics*[clip,width=40mm]{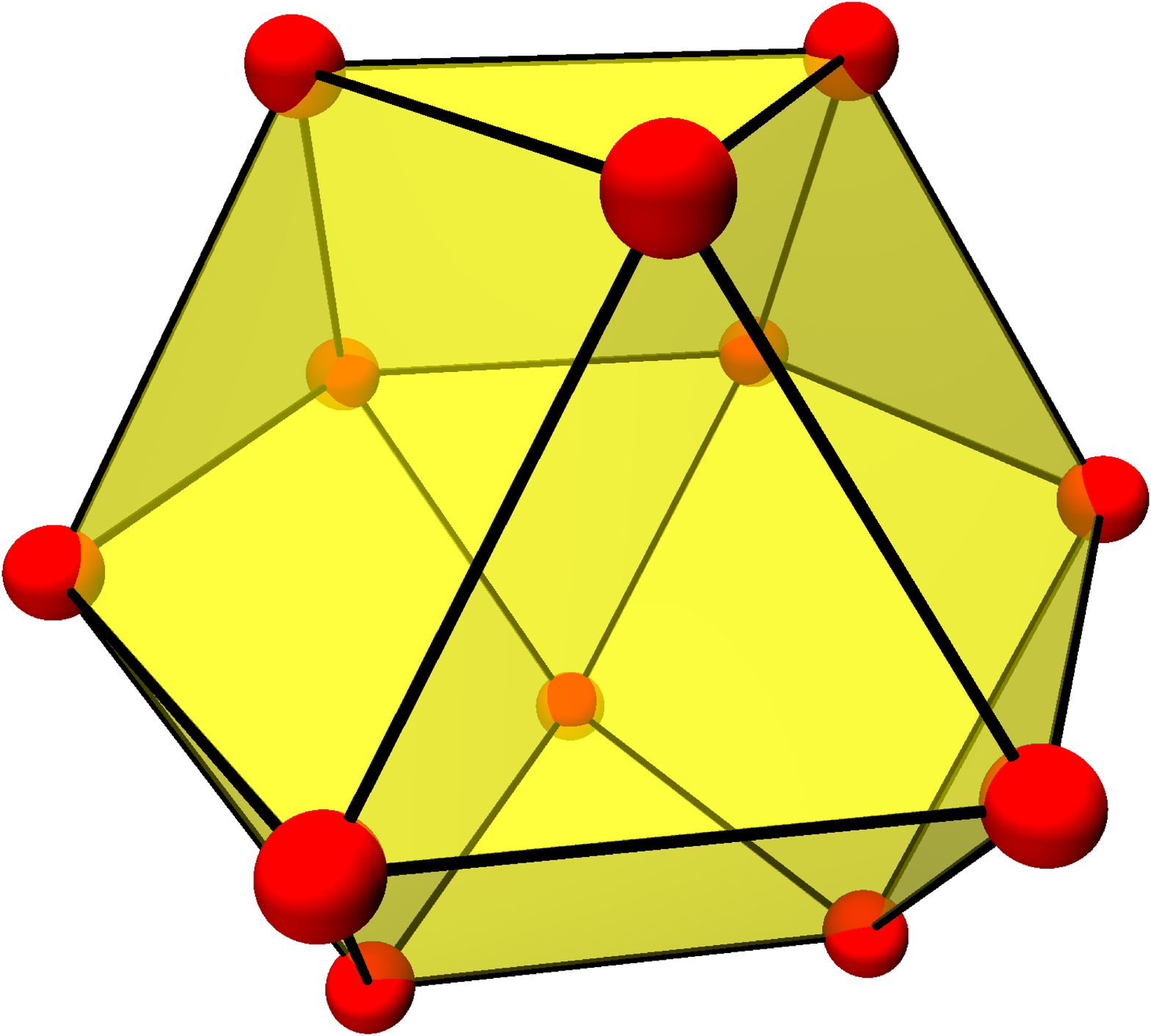}
\includegraphics*[clip,width=35mm]{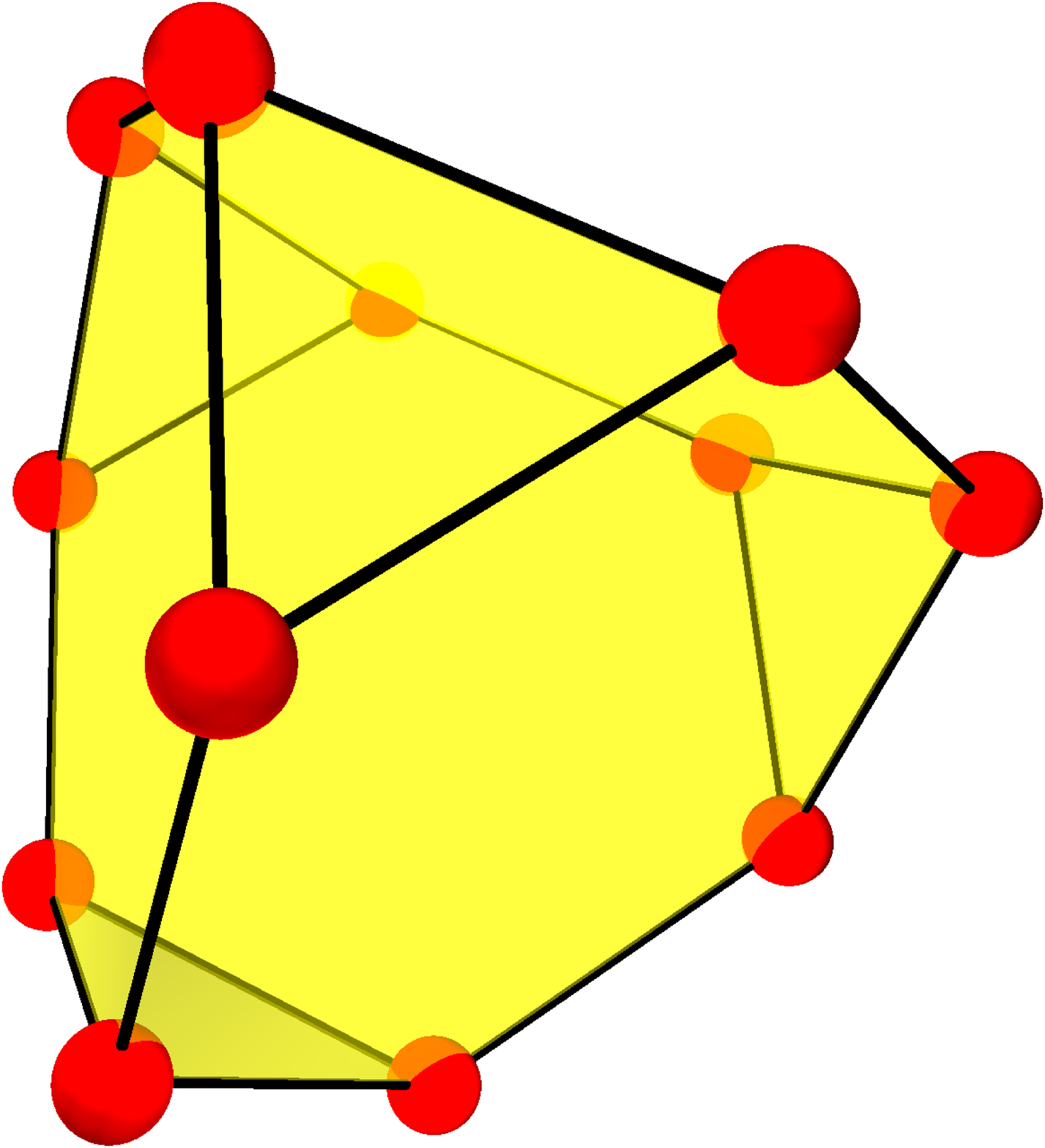}
\includegraphics*[clip,width=37mm]{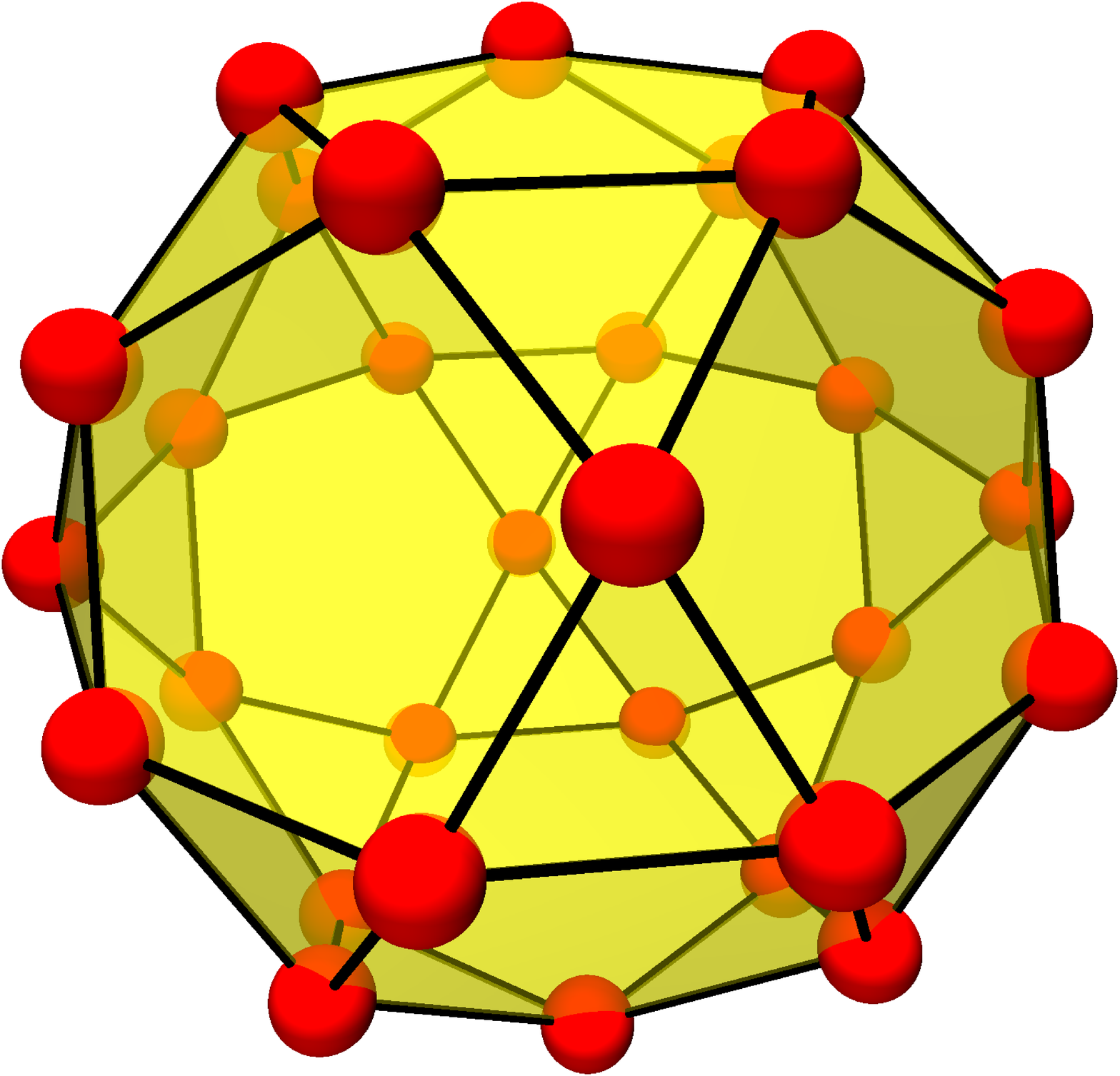}
\caption{Structure of the cuboctahedron (left), the truncated
  tetrahedron (middle), and the icosidodecahedron
  (right).}
\label{dt-schnack-F-4}
\end{figure}

Experience collected in condensed matter physics demonstrates
that prominent frustration effects can be observed in certain
spin lattices with antiferromagnetic nearest neighbor coupling
with the kagom\'{e} and the pyrochlore lattices being the most
prominent ones. Molecular realizations that come close to the
structure of the kagom\'{e} lattice do exist in the form of
cuboctahedra and
icosidodecahedra\cite{MSS:ACIE99,MLS:CPC01,AxL:PRB01,NTF:JPCS04,BKH:CC05,TMB:ACIE07}.
These systems, as can be seen in Fig.~\ref{dt-schnack-F-4}, consist
of corner sharing triangles. There are of course many other
frustrated molecules, among them the truncated
tetrahedra\cite{PLK:CC07}, and (unfortunately chemically
incomplete) icosahedra\cite{TEM:CEJ06}. Not every one of them
shows all of the below discussed properties.

\subsection{Ground state degeneracy}

An important effect of frustration can be given by a degenerate
ground state. Olivier Kahn discusses this degeneracy for
antiferromagnetic triangles with half-integer spins. Besides the
trivial $M$-degeneracy of the $S=1/2$ ground state one observes
a twofold degeneracy, which in the language of cyclic groups is
due to the symmetry $k \Leftrightarrow (N-k)$, compare
eq.~\fmref{E-2-3}. The ground state of the respective systems
with integer spins is non-degenerate, and thus does not show
this frustration effect\cite{Kahn:CPL97}. Nevertheless, even in
the absence of ground state degeneracy there can be other
frustration effects as we will see later.

More generally one can state that all symmetric systems of
half-integer spins which can be described by trigonal point
groups are characterized by doublet ground states\cite{ZTT:IC07}
which belong to the irreproducible representation $^2E$.  A
physical consequence of a degenerate ground state is that these
spin systems are very likely to be perturbed by small
interactions not present in the Hamiltonian \fmref{E-2-1}. Such
interactions are given by structural, e.g. spontaneous
Jahn-Teller distortions, antisymmetric
exchange\cite{TTM:JMS07,KTM:DT10} or spin-Peierls
transitions\cite{RDS:PRL04}.

\subsection{Spin rings}

A natural extension of triangles is given by odd-membered spin
rings. In contrast to a large number of synthesized
even-membered spin rings such as ferric or chromium
wheels\cite{TDP:JACS94,ACC:ICA00,WKS:IO01,SSG:CAEJ02,LOP:ACIE03},
odd-membered rings are difficult to synthesize due to steric
hindrance of the ligands. Nevertheless, thanks to synthetic
cleverness several odd-membered antiferromagnetic rings exist,
although not quite as symmetric as their even-membered
counterparts\cite{LMM:03,CGS:ACIE04,CGS:JMMM05,YWM:CC06,FKK:PRB09,HNN:JACS09}.
In the classical ground state of a perfect odd-membered ring
adjacent spins are no longer aligned antiparallel, their
orientation can for instance be described as non-collinear,
i.e. a M{\"o}bius strip\cite{CGS:ACIE04,CGS:JMMM05}.

\begin{figure}[ht!]
\centering
\includegraphics[clip,width=75mm]{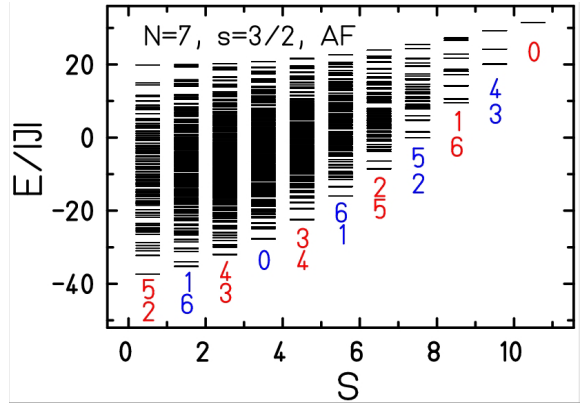}
\caption{Energy eigenvalues of a spin ring ($N=7, s=3/2$) with
  antiferromagnetic nearest neighbor coupling as function of
  total spin $S$.}
\label{dt-schnack-F-5}
\end{figure}

Figure~\ref{dt-schnack-F-5} shows the energy spectrum of a spin
ring ($N=7, s=3/2$) with antiferromagnetic nearest neighbor
coupling as function of total spin $S$. The numbers close to the
lowest levels for each $S$ denote again the momentum quantum
numbers of this relative ground state. In contrast to bipartite
rings one sees that many of them are twofold degenerate (in
addition to their $M$-degeneracy). This is a consequence of the
symmetry $k \Leftrightarrow (N-k)$ which only for $k=0$ allows a
non-degenerate relative ground state. $k=N/2$ is impossible for
odd-membered rings.

There is striking numerical evidence that the momentum quantum
numbers $k$ for odd-membered rings are not random, but follow
certain rules as they do for even-membered
rings\cite{BSS:JMMM00:B}. It seems that one can even formulate a
generalized rule that holds both for even-membered (bipartite)
as well as for odd-membered (frustrated) rings,
\begin{eqnarray}
\label{E-3-1}
k
\equiv
\pm
a
\left\lceil
\frac{N}{2}
\right\rceil
\mod N
\ ,\quad
a=Ns-M
\ .
\end{eqnarray}
Here $\lceil N/2\rceil$ denotes the smallest integer greater
than or equal to $N/2$. It is interesting to see, that $k$ is
independent of $s$ for a given $N$ and $a$. This means that the
sequence of $k$ values starting at the largest $S$ value,
i.e. for $a=0$, is the same for rings of the same size $N$ but
with different single spins $s$.  Table~\ref{T-1} provides an
example. In addition it was found that for all $N$ except three,
the degeneracy is minimal, i.e. completely given by the
$M$-degeneracy and the $k$-degeneracy\cite{BHS:PRB03}.

\begin{table}[ht!]
\centering
\begin{tabular}{|c|c||l|l|l|l|l|l|l|}
\hline
 & &\multicolumn{7}{c|}{$a$\phantom{$\op{\vec{H}}$}}\\
$N$&$s$& 0& 1&2&3&4&5&\dots\\[1mm]
\hline
{9} & {1/2} & {0} & {$5\equiv 4$}  & {$10\equiv 1$} & {$15\equiv 3$} & {$20\equiv 2$} & - & - \phantom{$\op{\vec{H}}$}  \\
\hline
{9} & {1}   & {0} & {$5\equiv 4$}  & {$10\equiv 1$} & {$15\equiv 3$} & {$20\equiv 2$} &
{$25\equiv 2$} & \dots \phantom{$\op{\vec{H}}$}  \\
\hline
\end{tabular}
\caption{Example of $k$ quantum numbers for an odd-membered ring
  with $N=9$ and $s=1/2$ as well as $s=1$. $20\equiv 2$ means
  that $a\lceil N/2\rceil$ is 20, which is equivalent to 2 when
  taking the modulus with respect to $N$. If $k\ne 0$, then
  $N-k$ is also ground state $k$ quantum number. One see that
  the $k$ numbers are the same for the two cases. The first
  sequence is only shown down to $a=4$, which constitutes
  already the $S=1/2$ ground state in this case.} 
\label{T-1}
\end{table}

Summarizing, the clear frustration effect for odd-membered rings
is the non-trivial degeneracy of relative ground state levels
that does not appear for even-membered rings.

I would like to remark that frustration of antiferromagnetic
even membered spin rings can also be introduced by an
antiferromagnetic next nearest neighbor interaction. Again,
depending on the ratio of the two interactions the ground state
can be non-collinear\cite{MZB:PRL04,DMR:PRL05}. However, it
could be demonstrated that the level ordering $E(S+1)>E(S)$ is
rather robust for small next nearest neighbor
interactions\cite{RIR:JMMM95,RIV:JLTP95}.

\subsection{Low-lying singlets}

The idea of a non-trivial ground state degeneracy is also used
in discussions of frustrated classical spin systems such as the
triangle shown in Fig.~\ref{dt-schnack-F-1}. The classical
ground state on the r.h.s. is two-fold non-trivially
(i.e. orbitally) degenerate because the two lower spins could be
exchanged to yield another ground state. The two ground states
differ in their chirality which expresses itself in different
signs of the $z$-component of the orbital (pseudo spin)
moment\cite{ZTT:IC07}.  The trivial degeneracy is given by
collective rotations of the full ground state. In several
classical systems, e.g. the kagom\'{e} lattice antiferromagnet,
the non-trivial degeneracy is actually infinity,
i.e. macroscopic, and this means it scales with the size of the
system. Nevertheless, for such systems one may find (or believe
or have numerical evidence) that the ground state of the
corresponding quantum system is non-degenerate. In the language
of higher-order spin wave theory one would express this
observation as that the classical degeneracy is lifted ``by
quantum fluctuations", a concept that sometimes is also denoted
as ``order from disorder" \cite{Hen:PRL89}. However, the
classically degenerate ground states do not move far, they
appear as low-lying singlets (below the first triplet) in the
quantum spectrum\cite{WEB:EPJB98,BAA:PRL03}.

\begin{figure}[ht!]
\centering
\includegraphics[clip,width=55mm]{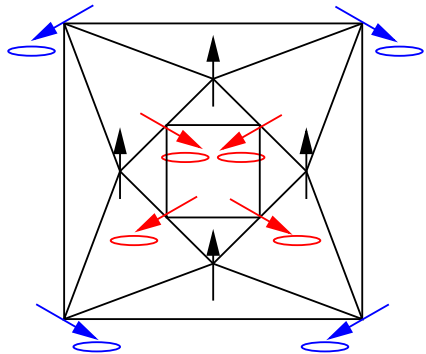}
\caption{Planar graph of the cuboctahedron: the classical spins
  reside at the vertices, the solid edges denote the
  antiferromagnetic exchange interactions. Two groups of
  classical spins (outer and inner square) can independendly be
  rotated without changing the ground state energy.}
\label{dt-schnack-F-6}
\end{figure}

Figure~\ref{dt-schnack-F-6} provides an interesting example of non-trivial
classical ground-state degeneracy that is related to the
kagom\'{e}. It shows the graph of antiferromagnetic interactions
in a cuboctahedron with spins at the vertices. The classical
ground state is given by states with relative angles of
120$^\circ$ between neighboring spins. There exist an infinite
number of non-trivially degenerate ground states that can be
produced by independent collective rotations of groups of four
spins (outer and inner square). The same holds for the
kagom\'{e} where the squares are just replaced by hexagons. A
quantum treatment lifts this degeneracy, and the quantum ground
state of the cuboctahedron is non-degenerate.

\begin{figure}[ht!]
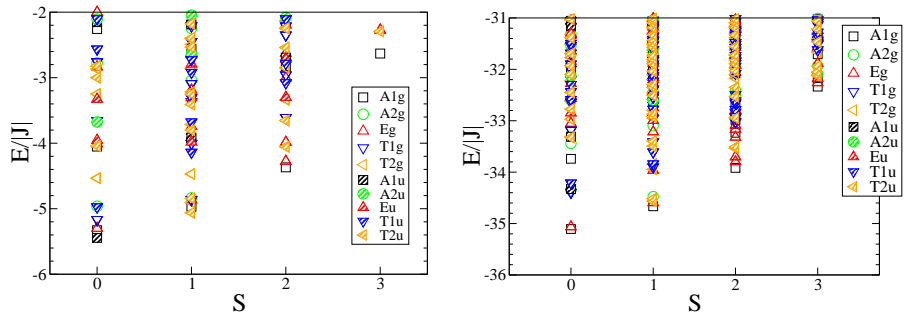

\centering
\includegraphics[clip,width=55mm]{dt-schnack-fig-7a.eps}
\quad
\includegraphics[clip,width=58mm]{dt-schnack-fig-7b.eps}
\caption{Low-lying energy spectrum of the antiferromagnetic
  cuboctahedron for $s=1/2$ (l.h.s.) and for $s=3/2$
  (r.h.s.). The levels are characterized by the irreducible
  representations of $O_h$.}
\label{dt-schnack-F-7}
\end{figure}

Figure~\ref{dt-schnack-F-7} shows the low-lying energy spectrum of the
antiferromagnetic cuboctahedron for $s=1/2$ (l.h.s.) and for
$s=3/2$ (r.h.s.). The ground states of both systems are
non-degenerate. One notices that several levels (in the case of
$s=1/2$) and one additional level (in the case of $s=3/2$) exist
below the first
triplet\cite{SSR:JMMM05,ScS:PRB09,ScS:P09,Schnalle:Diss09}. Such a behavior
is also expected for molecules of icosidodecahedral structure
and numerically evaluated for the case of $s=1/2$ which would
correspond to the vanadium Keplerate\cite{SSR:JMMM05,RLM:PRB08}. 

Although the low-lying singlets are magnetically silent they
have a great impact on the specific heat which at low
temperatures may thus exhibit extra features. For the
cuboctahedron for $s=1/2$ for instance one observes a sharp
additional peak at low temperatures\cite{SSR:JMMM05}. The
magnetic susceptibility is only mildly influenced through the
partition function. Numerical studies suggest that the number of
low-lying singlets decreases with increasing number of the
individual spin quantum number of the
system\cite{SSR:JMMM05,ScS:P09}.

\subsection{Magnetization plateaus}

Regular, i.e. highly symmetric, bipartite finite
antiferromagnets possess a low-temperature magnetization curve
that has the form of a regular staircase. Frustrated systems may
exhibit unusual deviations thereof. Such deviations appear more
pronounced in infinitely extended quantum antiferromagnets as
again the kagom\'{e} and pyrochlore lattice since there even the
$(T=0)$ magnetization curve is a continuous (even
differentiable) curve except for plateaus and jumps.

\begin{figure}[ht!]
\centering
\includegraphics[clip,width=55mm]{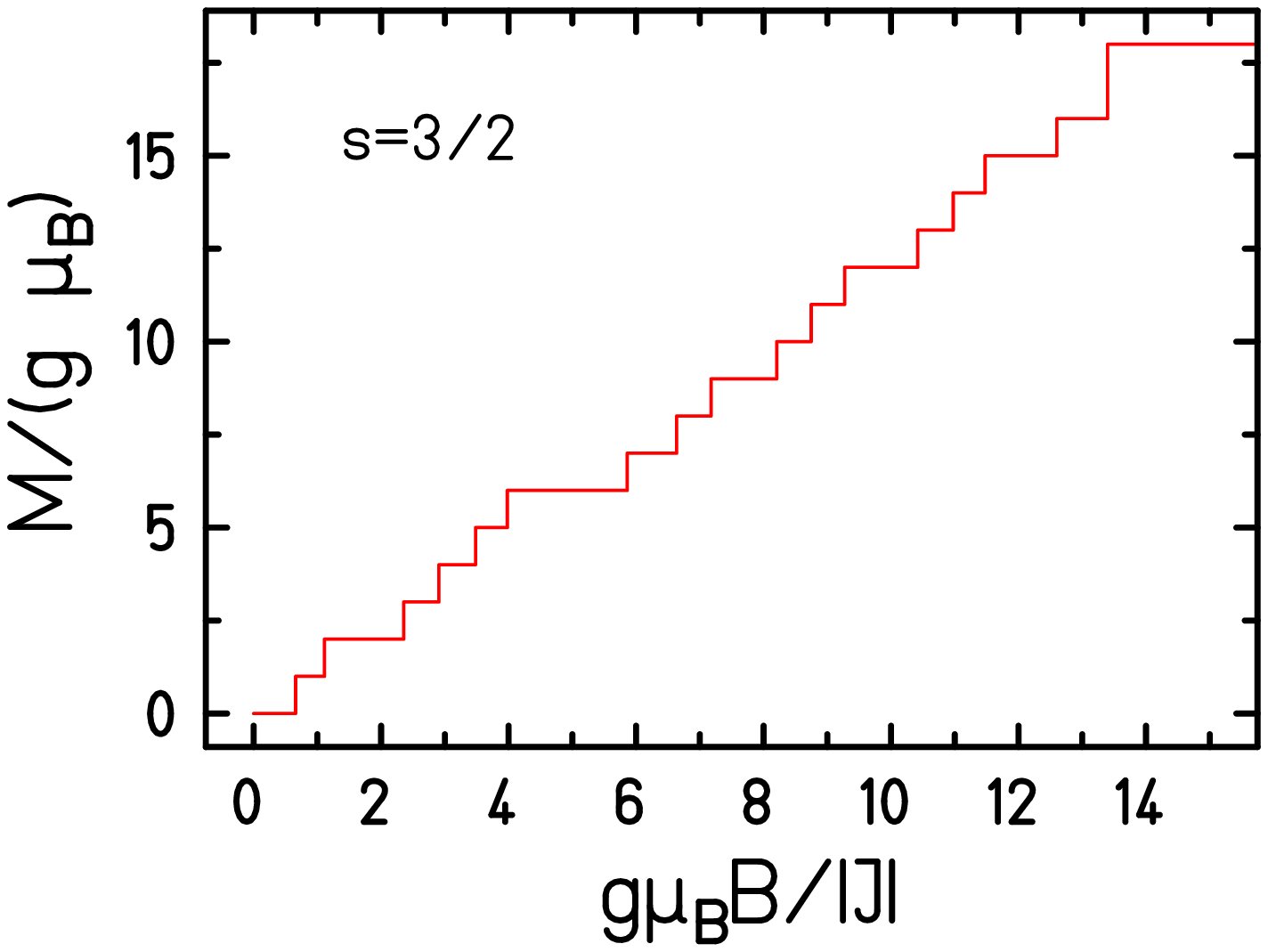}
\quad
\includegraphics[clip,width=55mm]{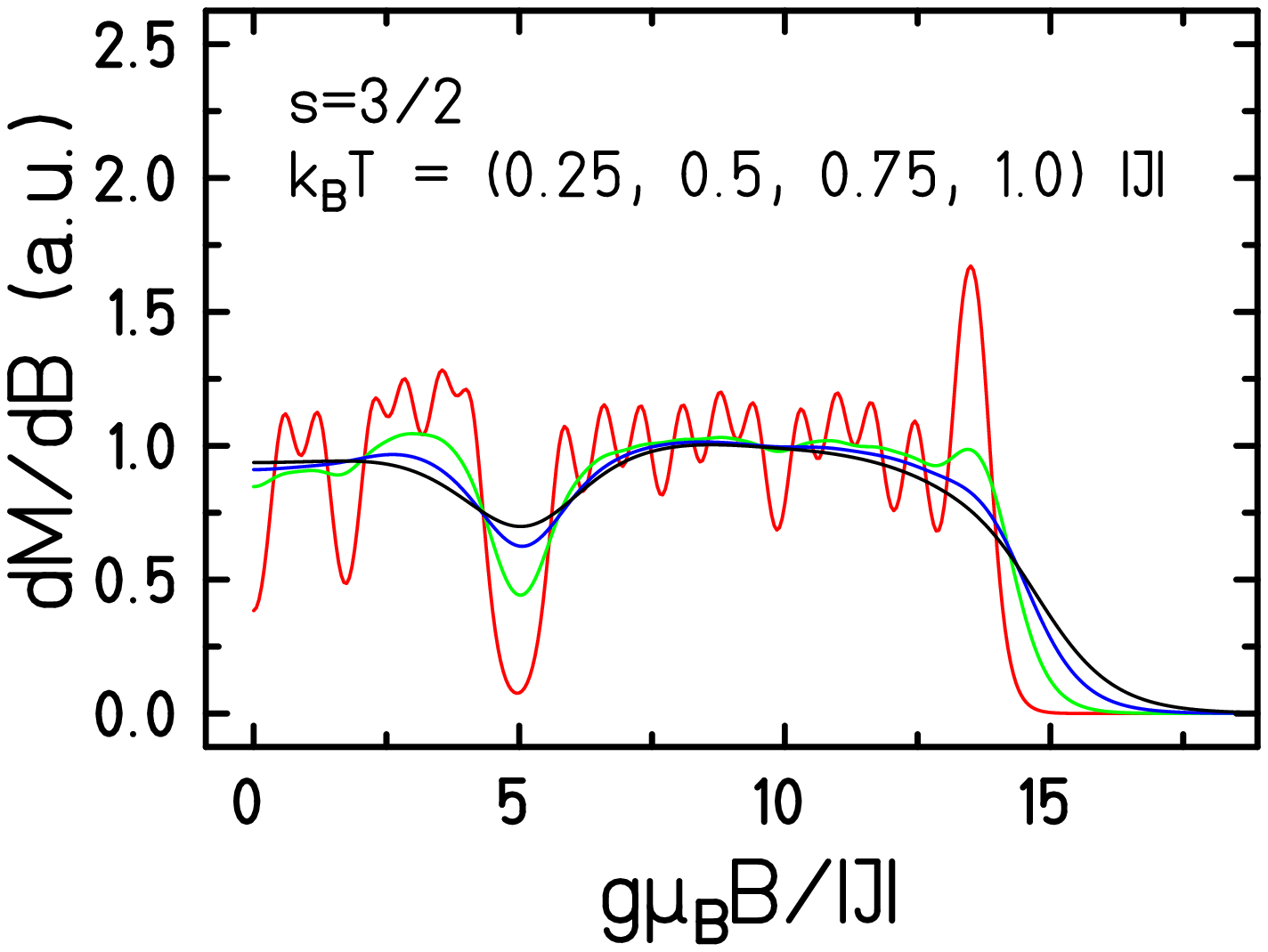}
\caption{L.h.s.: Magnetization curve at $T=0$ for the
  antiferromagnetic cuboctahedron with $s=3/2$. R.h.s.:
  Differential susceptibility for the same system for several
  small temperatures. The higher the temperature the smoother
  the corresponding curve. At temperatures $k_B T\geq 0.75 |J|$
  only one feature persists, the dip at $g\mu_BB/|J|\approx 5$
  which is approximately one third of the saturation field.}
\label{dt-schnack-F-8}
\end{figure}

In finite antiferromagnets the magnetization curve at $T=0$ is a
(non-differentiable) staircase anyhow due to successive level
crossings. Nevertheless, unusual plateaus can be
identified. Figure~\ref{dt-schnack-F-8} shows the magnetization curve at
$T=0$ for the antiferromagnetic cuboctahedron with $s=3/2$. One
can clearly see that the plateau at $1/3$ of the saturation
magnetization is wider than the others. The differential
susceptibility on the r.h.s. demonstrates that it is thermally
also much more stable since it leads to a magnetization dip even
at those temperatures where the features stemming from the other
steps of the staircase have already disappeared\cite{ScS:P09}.

The plateau at $1/3$ of the saturation magnetization appears in
systems built of corner sharing triangles such as the kagom\'{e}
lattice
antiferromagnet\cite{AHL:PRL98,Atw:NM02,Zhi:PRL02,HPZ:PB02,CGH:PRB05}
or molecular realizations such as the cuboctahedron and the
icosidodecahedron\cite{SSR:JMMM05,SNS:PRL05,RLM:PRB08}. Its
stability is classically related to the dominating contribution
of so-called ``up-up-down'' ({\it uud}) spin configurations to
the partition function\cite{KaM:JPSJ85,CGH:PRB05}. Recently it
could be shown for the cuboctahedron and the icosidodecahedron
that the corresponding quantum states indeed also dominate the
quantum partition function\cite{RLM:PRB08}. It is worth
mentioning that the differential susceptibility of a simple
quantum mechanical triangle (both for integer and half-integer
spins) also shows the dip at $1/3$ for non-zero temperatures
although it does not exhibit the plateau at $T=0$. The reason is
given by the special degeneracy of energy levels for larger
$S$. This is the point where a spin triangle of integer spins
with a boring $^1A_1$ ground state shows its frustration
effects.

For the pyrochlore the argumentation is analogous with the
difference that in this case a half-magnetization plateau is
stabilized by ``up-up-up-down'' ({\it uuud})
configurations\cite{PSS:PRL04}.

\subsection{Magnetization jumps}

\begin{figure}[ht!]
\centering
\includegraphics[clip,width=105mm]{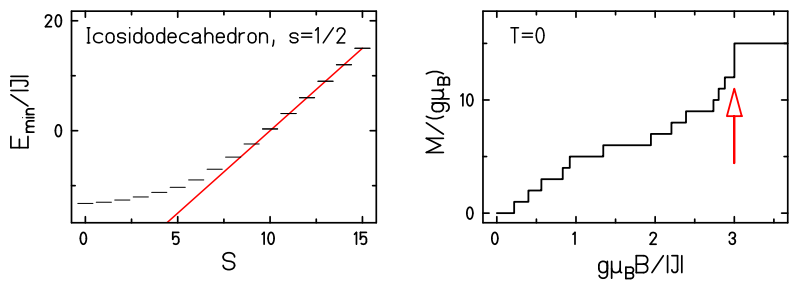}
\caption{Minimal energies of an antiferromagnetic
  icosidodecahedron with $s=1/2$ (l.h.s.) and resulting
  magnetization curve at $T=0$ (r.h.s.).}
\label{dt-schnack-F-9}
\end{figure}

Many observable effects of frustration had their first
(theoretical) discovery in condensed matter physics of infinite
lattices. But a certain class of unusual magnetization jumps was
discovered in connection with the antiferromagnetic Keplerate
molecules of icosidodecahedral
structure\cite{SSR:EPJB01}. Figure~\ref{dt-schnack-F-9} displays on the
l.h.s. the minimal energies of an antiferromagnetic
icosidodecahedron with $s=1/2$. The highest four levels follow a
strict linear dependence (highlighted by the straight line),
which results in a magnetization jump to saturation of $\Delta
{\mathcal M}/(g\mu_B)=3$, compare r.h.s. of
Fig.~\ref{dt-schnack-F-9}. Jumps of non-trivial height, i.e. with $\Delta
{\mathcal M}/(g\mu_B)>1$, will always occur if the curve of
minimal energy levels as function of total spin is not convex,
i.e. linear or concave\cite{LhM:02}. Antiferromagnetic clusters
with a rotational-band of minimal energies, that is thus
automatically convex, will therefore never exhibit such jumps.

In the case of the special jumps observed in the
icosidodecahedron it turned out that an analytical model could
be devised that describes the many-body states of minimal energy
close to the highest energy, i.e. those connected by a straight
line on the l.h.s. of Fig.~\ref{dt-schnack-F-9} in terms of new
quasi-particles, independent localized
magnons\cite{SHS:PRL02,RSH:JPCM04}. Such states govern the
low-temperature high-field behavior of many frustrated
antiferromagnets such as the kagom\'{e} and the pyrochlore
lattice again, but also several other spin structures as for
instance special saw-tooth chains like
azurite\cite{ZhT:IKYS05,SBD:PRB06,RWS:PRL08}.  The number of
these special many-body states does not depend on the individual
spin quantum number but on the topological structure of
interactions. It may easily grow exponentially with the size of
the system, as for lattices\cite{DeR:PRB04,SRM:JPA06,ZhT:PRB07}. Due to
the dependence on the graph of interactions similar states do
exist for the Hubbard model\cite{DHR:PRB07}, where they lead to
flat-band ferromagnetism, a phenomenon which was discovered
already 15 years
ago\cite{Tas:PRL92,Mie:JPA92A,Mie:JPA92B,MiT:CMP93}.

The magnetization jump at the saturation field is a real quantum
phenomenon; it is connected to the simultaneous crossing of many
Zeeman levels. This means that even at $T=0$ such systems
possess non-zero
entropy\cite{ZhH:JSM04,SSR:PRB07,HoZ:JPCS09}. Therefore, in the
vicinity of the saturation field one can observe strong
magnetocaloric effects. Isentropes (curves of constant entropy)
exhibit steep and opposite slopes on either sides of the
saturation field, i.e. it would be possible to heat or cool the
system by just varying the external magnetic field. This was
also predicted for the classical counterparts\cite{Zhi:PRB03}.
Complementary to this high-field magnetocaloric effect there is
of course the possibility to try to achieve a huge ground state
spin which would lead to a pronounced zero-field magnetocaloric
effect\cite{MSS:JMMM92,BMT:JAP94,AGC:APL04,ECG:APL05,evangelisti:104414}.

\begin{figure}[ht!]
\centering
\includegraphics[clip,width=55mm]{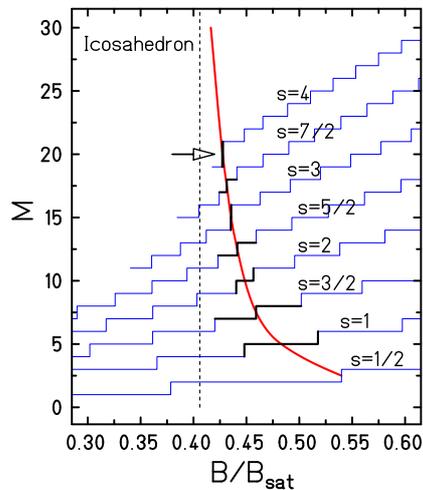}
\caption{Parts of the $T=0$ magnetization curves for
  antiferromagnetic icosahedra with $s=1/2, 1, 3/2, 2, 5/2, 3,
  7/2, 4$ (from bottom to top).}
\label{dt-schnack-F-10}
\end{figure}

Frustrated antiferromagnetic lattices of corner-sharing
triangles or tetrahedra are well investigated, systems of
edge-sharing polygons or bodies -- except for the triangular
lattice antiferromagnet -- not so much. In the case of molecules
this would refer for instance to the dodecahedron or the
icosahedron, the latter being a rather natural structure for
instance in cluster physics, but difficult to realize in
supramolecular chemistry\cite{TEM:CEJ06}.  It was theoretically
found that these special systems show another frustration
effect, again a jump, but not at saturation. Such a transition
is sometimes called metamagnetic phase
transition\cite{CoT:PRL92,SSS:PRL05,Kon:PRB07} since it
constitutes a so-called $(T=0)$-phase transition where one can
switch with the help of the external magnetic field between the
phases left and right of the jump. In classical investigations
the switching is accompanied by metastability and a
hysteresis\cite{SSS:PRL05}.  Figure~\ref{dt-schnack-F-10} displays parts of
$(T=0)$-magnetization curves for various single spin quantum
numbers $s$. Since the unusual magnetization jump is most
pronounced for classical spin icosahedra it is expected to show
up for larger $s$. As can be see in the figure a jump of twice
the normal height occurs for $s=4$, but for the smaller spins
shrinking magnetization steps act as precursors.

\subsection{Reduction of local moments}

\begin{figure}[ht!]
\centering
\includegraphics[clip,width=45mm]{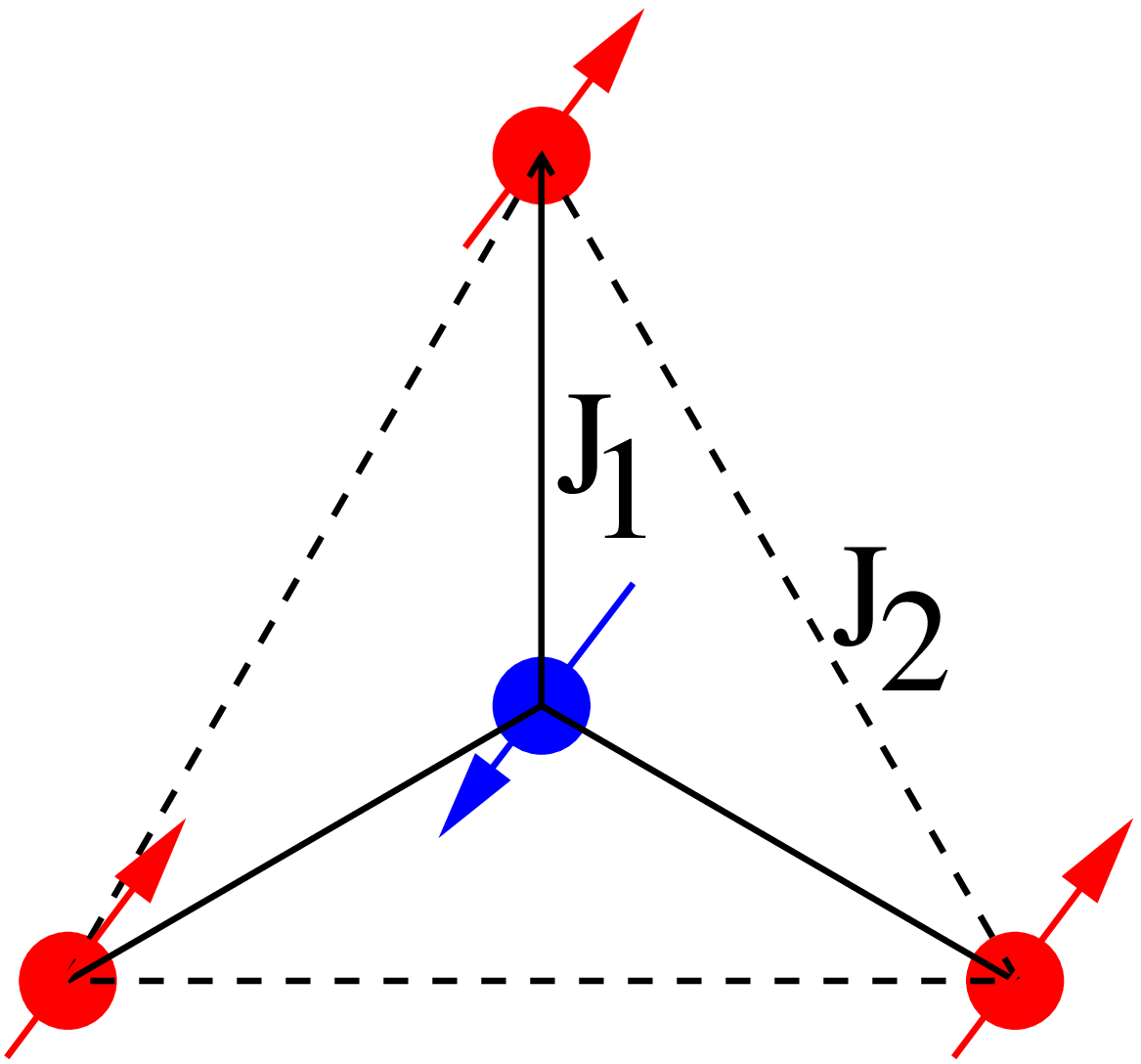}
\quad
\includegraphics[clip,width=65mm]{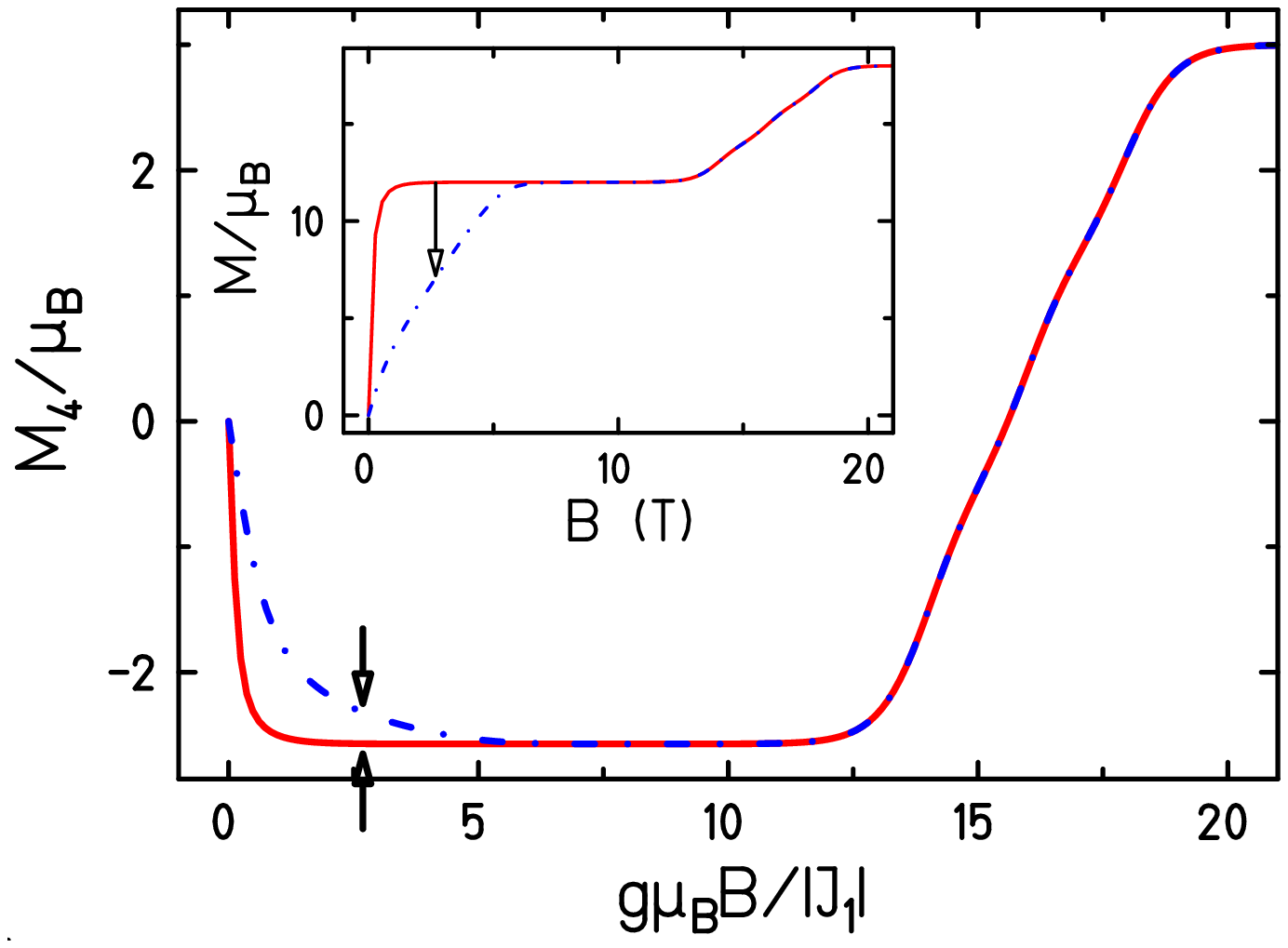}
\caption{L.h.s.: Sketch of a fictitious magnetic molecule with
  two exchange pathways. R.h.s.: Local magnetization for the
  central spin at $T=0.5|J_1|$
  for $J_2=0$ (solid curve) and $J_2=0.5 J_1$ (dashed dotted
  curve). The inset shows the total magnetization and the main
  figure the contribution of the central spin.}
\label{dt-schnack-F-11}
\end{figure}

As a last example I would like to discuss the influence of a
frustrating interaction on local spin moments. Such moments can
nowadays be measured in order to better understand the internal
magnetization distribution and thus the low-energy wave
functions. Experimental probes are for instance Nuclear Magnetic
Resonance\cite{MFK:PRL06} or -- and then even element-selective
-- X-ray magnetic circular dichroism\cite{CMB:PRB08}.  I would
like to use a fictitious star-like molecule for this discussion,
compare l.h.s. of Fig.~\ref{dt-schnack-F-11}, that is similar to recently
synthesized and investigated examples\cite{SSB:DT06,KBG:DT07,PKT:09,TBS:DT09}.

Figure~\ref{dt-schnack-F-11} shows on the r.h.s. the magnetization of the
fictitious molecule with three spins $s=5/2$ in the outer
triangle and a central spin with $s_4=3/2$. The coupling $J_1$
is antiferromagnetic. For $J_2=0$ the system is bipartite and
the ground state spin according to Lieb, Schultz, and Mattis is
at most ${\mathcal S}=|S_A-S_B|=3\times 5/2-3/2=6$. That this is
indeed the case can be inferred from the solid magnetization
curve in the inset, where one can see that the magnetization
immediately follows a Brillouin function of a total spin $S=6$
for small magnetic fields. In addition the main graphics shows
the magnetization contribution of the central spin which at low
temperatures ($T=0.5|J_1|$ in the example) and low-fields points
opposite to the three outer spins with a practically maximal
amplitude of ${\mathcal M}_4\approx 3\mu_B$.

With increasing frustration due to the antiferromagnetic
interaction $J_2$ the properties
change. Figure~\ref{dt-schnack-F-11} shows as a second example
the case of $J_2=0.5 J_1$ for the same temperature. The
dashed-dotted curve in the inset displays how much the total
magnetization shrinks especially at low field values. The arrow
depicts an example case. As can be seen in the main graphics the
contribution of the central spin also changes but not too much,
which leads to the conclusion that the frustration more strongly
modifies the joint moment of the three outer spins. This is a
special example of the more general phenomenon of reduction of
correlations $\langle \, \op{\vec{s}}_i \cdot \op{\vec{s}}_j \,
\rangle$ between spins due to frustration\cite{Schnack:PRB00}.

\subsection{Outlook}

Although the article describes the major frustration effects for
antiferromagnetic spin systems they can of course be discussed
in more depth. For the interested reader I would like to
recommend three specialized
books\cite{SRF:LNP04,Diep05,LMM:10}. Nowadays, research
interests in condensed matter physics focus on e.g. quantum
(i.e. $T=0$) phase transitions driven by frustration as for
instance in antiferromagnetic spin chains with antiferromagnetic
next nearest neighbor coupling or on exotic behavior as spin
liquid or spin
ice\cite{MBL:PRL98,JPSJ.78.093701,0295-5075-81-1-17006,BrG:Science01}.
The modern language is a description in terms of
quasi-particles. For example, the frustration effects in spin
chains with nearest and next nearest antiferromagnetic
interactions are denoted as ``condensation of triplets". The
afore discussed magnetization jumps in the cuboctahedron,
icosidodecahedron, kagom\'{e}, or pyrochlore can be well
understood by introducing localized independent magnons as quasi
particles. How far these concepts go and how real the
quasi-particles are demonstrates the recent experimental
verification of ``magnetic monopoles" in spin
ice\cite{BGC:Nature09}. 

Although magnetic molecules are only of finite size
(zero-dimensional) many of the exciting properties carry over as
I hope I have shown.

\section*{Acknowledgment}

This work was supported by the German Science Foundation (DFG)
through the research group 945. I would also like to thank Roman
Schnalle and Christian Karlewski for producing some of the
figures for me as well as Roman Schnalle, Stephen Blundell
(Oxford), Johannes Richter (Magdeburg) and Boris Tsukerblat (Be'er
Sheva) for carefully reading the manuscript. Johannes Richter
suggested to show Fig.~\ref{dt-schnack-F-6}.


\begin{mcitethebibliography}{102}
\providecommand*{\natexlab}[1]{#1}
\providecommand*{\mciteSetBstSublistMode}[1]{}
\providecommand*{\mciteSetBstMaxWidthForm}[2]{}
\providecommand*{\mciteBstWouldAddEndPuncttrue}
  {\def\EndOfBibitem{\unskip.}}
\providecommand*{\mciteBstWouldAddEndPunctfalse}
  {\let\EndOfBibitem\relax}
\providecommand*{\mciteSetBstMidEndSepPunct}[3]{}
\providecommand*{\mciteSetBstSublistLabelBeginEnd}[3]{}
\providecommand*{\EndOfBibitem}{}
\mciteSetBstSublistMode{f}
\mciteSetBstMaxWidthForm{subitem}
{(\emph{\alph{mcitesubitemcount}})}
\mciteSetBstSublistLabelBeginEnd{\mcitemaxwidthsubitemform\space}
{\relax}{\relax}

\bibitem[Kahn(1997)]{Kahn:CPL97}
O.~Kahn, \emph{Chem. Phys. Lett.}, 1997, \textbf{265}, 109--114\relax
\mciteBstWouldAddEndPuncttrue
\mciteSetBstMidEndSepPunct{\mcitedefaultmidpunct}
{\mcitedefaultendpunct}{\mcitedefaultseppunct}\relax
\EndOfBibitem
\bibitem[Toulouse(1977)]{Tou:CP77}
G.~Toulouse, \emph{Comm. Phys.}, 1977, \textbf{2}, 115--119\relax
\mciteBstWouldAddEndPuncttrue
\mciteSetBstMidEndSepPunct{\mcitedefaultmidpunct}
{\mcitedefaultendpunct}{\mcitedefaultseppunct}\relax
\EndOfBibitem
\bibitem[Kirkpatrick(1977)]{Kir:PRB77}
S.~Kirkpatrick, \emph{Phys. Rev. B}, 1977, \textbf{16}, 4630--4641\relax
\mciteBstWouldAddEndPuncttrue
\mciteSetBstMidEndSepPunct{\mcitedefaultmidpunct}
{\mcitedefaultendpunct}{\mcitedefaultseppunct}\relax
\EndOfBibitem
\bibitem[Ramirez(1994)]{Ram:ARMS94}
A.~P. Ramirez, \emph{Annu. Rev. Mater. Sci.}, 1994, \textbf{24}, 453\relax
\mciteBstWouldAddEndPuncttrue
\mciteSetBstMidEndSepPunct{\mcitedefaultmidpunct}
{\mcitedefaultendpunct}{\mcitedefaultseppunct}\relax
\EndOfBibitem
\bibitem[Greedan(2001)]{Gre:JMC01}
J.~Greedan, \emph{J. Mater. Chem.}, 2001, \textbf{11}, 37\relax
\mciteBstWouldAddEndPuncttrue
\mciteSetBstMidEndSepPunct{\mcitedefaultmidpunct}
{\mcitedefaultendpunct}{\mcitedefaultseppunct}\relax
\EndOfBibitem
\bibitem[Moessner(2001)]{Moe:CJP01}
R.~Moessner, \emph{Can. J. Phys.}, 2001, \textbf{79}, 1283\relax
\mciteBstWouldAddEndPuncttrue
\mciteSetBstMidEndSepPunct{\mcitedefaultmidpunct}
{\mcitedefaultendpunct}{\mcitedefaultseppunct}\relax
\EndOfBibitem
\bibitem[Bramwell and Gingras(2001)]{BrG:Science01}
S.~T. Bramwell and M.~J.~P. Gingras, \emph{Science}, 2001, \textbf{294},
  1495--1501\relax
\mciteBstWouldAddEndPuncttrue
\mciteSetBstMidEndSepPunct{\mcitedefaultmidpunct}
{\mcitedefaultendpunct}{\mcitedefaultseppunct}\relax
\EndOfBibitem
\bibitem[Lieb \emph{et~al.}(1961)Lieb, Schultz, and Mattis]{LSM:AP61}
E.~H. Lieb, T.~Schultz and D.~C. Mattis, \emph{Ann. Phys. (N.Y.)}, 1961,
  \textbf{16}, 407\relax
\mciteBstWouldAddEndPuncttrue
\mciteSetBstMidEndSepPunct{\mcitedefaultmidpunct}
{\mcitedefaultendpunct}{\mcitedefaultseppunct}\relax
\EndOfBibitem
\bibitem[Lieb and Mattis(1962)]{LiM:JMP62}
E.~H. Lieb and D.~C. Mattis, \emph{J.~Math. Phys.}, 1962, \textbf{3}, 749\relax
\mciteBstWouldAddEndPuncttrue
\mciteSetBstMidEndSepPunct{\mcitedefaultmidpunct}
{\mcitedefaultendpunct}{\mcitedefaultseppunct}\relax
\EndOfBibitem
\bibitem[Marshall(1955)]{Mar:PRS55}
W.~Marshall, \emph{Proc. Royal. Soc. A (London)}, 1955, \textbf{232}, 48\relax
\mciteBstWouldAddEndPuncttrue
\mciteSetBstMidEndSepPunct{\mcitedefaultmidpunct}
{\mcitedefaultendpunct}{\mcitedefaultseppunct}\relax
\EndOfBibitem
\bibitem[Anderson(1952)]{And:PR52}
P.~W. Anderson, \emph{Phys. Rev.}, 1952, \textbf{86}, 694--701\relax
\mciteBstWouldAddEndPuncttrue
\mciteSetBstMidEndSepPunct{\mcitedefaultmidpunct}
{\mcitedefaultendpunct}{\mcitedefaultseppunct}\relax
\EndOfBibitem
\bibitem[Julien \emph{et~al.}(1999)Julien, Jang, Lascialfari, Borsa,
  Horvati\'c, Caneschi, and Gatteschi]{JJL:PRL99}
M.-H. Julien, Z.~Jang, A.~Lascialfari, F.~Borsa, M.~Horvati\'c, A.~Caneschi and
  D.~Gatteschi, \emph{Phys. Rev. Lett.}, 1999, \textbf{83}, 227\relax
\mciteBstWouldAddEndPuncttrue
\mciteSetBstMidEndSepPunct{\mcitedefaultmidpunct}
{\mcitedefaultendpunct}{\mcitedefaultseppunct}\relax
\EndOfBibitem
\bibitem[Schnack and Luban(2000)]{ScL:PRB00}
J.~Schnack and M.~Luban, \emph{Phys. Rev. B}, 2000, \textbf{63}, 014418\relax
\mciteBstWouldAddEndPuncttrue
\mciteSetBstMidEndSepPunct{\mcitedefaultmidpunct}
{\mcitedefaultendpunct}{\mcitedefaultseppunct}\relax
\EndOfBibitem
\bibitem[Abbati \emph{et~al.}(2000)Abbati, Caneschi, Cornia, Fabretti, and
  Gatteschi]{ACC:ICA00}
G.~L. Abbati, A.~Caneschi, A.~Cornia, A.~C. Fabretti and D.~Gatteschi,
  \emph{Inorg. Chim. Acta}, 2000, \textbf{297}, 291\relax
\mciteBstWouldAddEndPuncttrue
\mciteSetBstMidEndSepPunct{\mcitedefaultmidpunct}
{\mcitedefaultendpunct}{\mcitedefaultseppunct}\relax
\EndOfBibitem
\bibitem[Waldmann(2001)]{Wal:PRB01}
O.~Waldmann, \emph{Phys. Rev. B}, 2001, \textbf{65}, 024424\relax
\mciteBstWouldAddEndPuncttrue
\mciteSetBstMidEndSepPunct{\mcitedefaultmidpunct}
{\mcitedefaultendpunct}{\mcitedefaultseppunct}\relax
\EndOfBibitem
\bibitem[Dai and Whangbo(2004)]{dai:672}
D.~Dai and M.-H. Whangbo, \emph{J. Chem. Phys.}, 2004, \textbf{121},
  672--680\relax
\mciteBstWouldAddEndPuncttrue
\mciteSetBstMidEndSepPunct{\mcitedefaultmidpunct}
{\mcitedefaultendpunct}{\mcitedefaultseppunct}\relax
\EndOfBibitem
\bibitem[M\"uller \emph{et~al.}(1999)M\"uller, Sarkar, Shah, B\"ogge,
  Schmidtmann, Sarkar, K\"ogerler, Hauptfleisch, Trautwein, and
  Sch\"unemann]{MSS:ACIE99}
A.~M\"uller, S.~Sarkar, S.~Q.~N. Shah, H.~B\"ogge, M.~Schmidtmann, S.~Sarkar,
  P.~K\"ogerler, B.~Hauptfleisch, A.~Trautwein and V.~Sch\"unemann,
  \emph{Angew. Chem. Int. Ed.}, 1999, \textbf{38}, 3238\relax
\mciteBstWouldAddEndPuncttrue
\mciteSetBstMidEndSepPunct{\mcitedefaultmidpunct}
{\mcitedefaultendpunct}{\mcitedefaultseppunct}\relax
\EndOfBibitem
\bibitem[M\"uller \emph{et~al.}(2001)M\"uller, Luban, Schr\"oder, Modler,
  K\"ogerler, Axenovich, Schnack, Canfield, Bud'ko, and Harrison]{MLS:CPC01}
A.~M\"uller, M.~Luban, C.~Schr\"oder, R.~Modler, P.~K\"ogerler, M.~Axenovich,
  J.~Schnack, P.~C. Canfield, S.~Bud'ko and N.~Harrison, \emph{Chem. Phys.
  Chem.}, 2001, \textbf{2}, 517\relax
\mciteBstWouldAddEndPuncttrue
\mciteSetBstMidEndSepPunct{\mcitedefaultmidpunct}
{\mcitedefaultendpunct}{\mcitedefaultseppunct}\relax
\EndOfBibitem
\bibitem[Axenovich and Luban(2001)]{AxL:PRB01}
M.~Axenovich and M.~Luban, \emph{Phys. Rev. B}, 2001, \textbf{63}, 100407\relax
\mciteBstWouldAddEndPuncttrue
\mciteSetBstMidEndSepPunct{\mcitedefaultmidpunct}
{\mcitedefaultendpunct}{\mcitedefaultseppunct}\relax
\EndOfBibitem
\bibitem[Kunisada \emph{et~al.}(2009)Kunisada, Takemura, and
  Fukumoto]{NTF:JPCS04}
N.~Kunisada, S.~Takemura and Y.~Fukumoto, \emph{J. Phys.: Conf. Ser.}, 2009,
  \textbf{145}, 012083 (4pp)\relax
\mciteBstWouldAddEndPuncttrue
\mciteSetBstMidEndSepPunct{\mcitedefaultmidpunct}
{\mcitedefaultendpunct}{\mcitedefaultseppunct}\relax
\EndOfBibitem
\bibitem[Botar \emph{et~al.}(2005)Botar, K{\"o}gerler, and Hill]{BKH:CC05}
B.~Botar, P.~K{\"o}gerler and C.~L. Hill, \emph{Chem. Commun.}, 2005,
  3138--3140\relax
\mciteBstWouldAddEndPuncttrue
\mciteSetBstMidEndSepPunct{\mcitedefaultmidpunct}
{\mcitedefaultendpunct}{\mcitedefaultseppunct}\relax
\EndOfBibitem
\bibitem[Todea \emph{et~al.}(2007)Todea, Merca, B{\"o}gge, van Slageren,
  Dressel, Engelhardt, Luban, Glaser, Henry, and M{\"u}ller]{TMB:ACIE07}
A.~M. Todea, A.~Merca, H.~B{\"o}gge, J.~van Slageren, M.~Dressel,
  L.~Engelhardt, M.~Luban, T.~Glaser, M.~Henry and A.~M{\"u}ller, \emph{Angew.
  Chem. Int. Ed.}, 2007, \textbf{46}, 6106--6110\relax
\mciteBstWouldAddEndPuncttrue
\mciteSetBstMidEndSepPunct{\mcitedefaultmidpunct}
{\mcitedefaultendpunct}{\mcitedefaultseppunct}\relax
\EndOfBibitem
\bibitem[Pradeep \emph{et~al.}(2007)Pradeep, Long, K{\"o}gerler, and
  Cronin]{PLK:CC07}
C.~P. Pradeep, D.-L. Long, P.~K{\"o}gerler and L.~Cronin, \emph{Chem. Commun.},
  2007,  4254--4256\relax
\mciteBstWouldAddEndPuncttrue
\mciteSetBstMidEndSepPunct{\mcitedefaultmidpunct}
{\mcitedefaultendpunct}{\mcitedefaultseppunct}\relax
\EndOfBibitem
\bibitem[Tolis \emph{et~al.}(2006)Tolis, Engelhardt, Mason, Rajaraman, Kindo,
  Luban, Matsuo, Nojiri, Raftery, Schr\"oder, Tuna, Wernsdorfer, and
  Winpenny]{TEM:CEJ06}
E.~I. Tolis, L.~P. Engelhardt, P.~V. Mason, G.~Rajaraman, K.~Kindo, M.~Luban,
  A.~Matsuo, H.~Nojiri, J.~Raftery, C.~Schr\"oder, G.~A. T.~F. Tuna,
  W.~Wernsdorfer and R.~E.~P. Winpenny, \emph{Chem. Eur. J.}, 2006,
  \textbf{12}, 8961--8968\relax
\mciteBstWouldAddEndPuncttrue
\mciteSetBstMidEndSepPunct{\mcitedefaultmidpunct}
{\mcitedefaultendpunct}{\mcitedefaultseppunct}\relax
\EndOfBibitem
\bibitem[Zamstein \emph{et~al.}(2007)Zamstein, Tarantul, and
  Tsukerblat]{ZTT:IC07}
N.~Zamstein, A.~Tarantul and B.~Tsukerblat, \emph{Inorg. Chem.}, 2007,
  \textbf{46}, {8851--8858}\relax
\mciteBstWouldAddEndPuncttrue
\mciteSetBstMidEndSepPunct{\mcitedefaultmidpunct}
{\mcitedefaultendpunct}{\mcitedefaultseppunct}\relax
\EndOfBibitem
\bibitem[Tsukerblat \emph{et~al.}({2007})Tsukerblat, Tarantul, and
  M{\"u}ller]{TTM:JMS07}
B.~Tsukerblat, A.~Tarantul and A.~M{\"u}ller, \emph{J. Mol. Struct.}, {2007},
  \textbf{{838}}, {124--132}\relax
\mciteBstWouldAddEndPuncttrue
\mciteSetBstMidEndSepPunct{\mcitedefaultmidpunct}
{\mcitedefaultendpunct}{\mcitedefaultseppunct}\relax
\EndOfBibitem
\bibitem[K{\"o}gerler \emph{et~al.}(2010)K{\"o}gerler, Tsukerblat, and
  M{\"u}ller]{KTM:DT10}
P.~K{\"o}gerler, B.~Tsukerblat and A.~M{\"u}ller, \emph{Dalton Trans.},
  2010\relax
\mciteBstWouldAddEndPuncttrue
\mciteSetBstMidEndSepPunct{\mcitedefaultmidpunct}
{\mcitedefaultendpunct}{\mcitedefaultseppunct}\relax
\EndOfBibitem
\bibitem[Richter \emph{et~al.}({2004})Richter, Derzhko, and
  Schulenburg]{RDS:PRL04}
J.~Richter, O.~Derzhko and J.~Schulenburg, \emph{Phys. Rev. Lett.}, {2004},
  \textbf{{93}}, 107206\relax
\mciteBstWouldAddEndPuncttrue
\mciteSetBstMidEndSepPunct{\mcitedefaultmidpunct}
{\mcitedefaultendpunct}{\mcitedefaultseppunct}\relax
\EndOfBibitem
\bibitem[Taft \emph{et~al.}(1994)Taft, Delfs, Papaefthymiou, Foner, Gatteschi,
  and Lippard]{TDP:JACS94}
K.~L. Taft, C.~D. Delfs, G.~C. Papaefthymiou, S.~Foner, D.~Gatteschi and S.~J.
  Lippard, \emph{J.~Am. Chem. Soc.}, 1994, \textbf{116}, 823\relax
\mciteBstWouldAddEndPuncttrue
\mciteSetBstMidEndSepPunct{\mcitedefaultmidpunct}
{\mcitedefaultendpunct}{\mcitedefaultseppunct}\relax
\EndOfBibitem
\bibitem[Waldmann \emph{et~al.}(2001)Waldmann, Koch, Schromm, Sch\"ulein,
  M\"uller, Bernt, Saalfrank, Hampel, and Balthes]{WKS:IO01}
O.~Waldmann, R.~Koch, S.~Schromm, J.~Sch\"ulein, P.~M\"uller, I.~Bernt, R.~W.
  Saalfrank, F.~Hampel and E.~Balthes, \emph{Inorg. Chem.}, 2001, \textbf{40},
  2986\relax
\mciteBstWouldAddEndPuncttrue
\mciteSetBstMidEndSepPunct{\mcitedefaultmidpunct}
{\mcitedefaultendpunct}{\mcitedefaultseppunct}\relax
\EndOfBibitem
\bibitem[van Slageren \emph{et~al.}({2002})van Slageren, Sessoli, Gatteschi,
  Smith, Helliwell, Winpenny, Cornia, Barra, Jansen, Rentschler, and
  Timco]{SSG:CAEJ02}
J.~van Slageren, R.~Sessoli, D.~Gatteschi, A.~Smith, M.~Helliwell, R.~Winpenny,
  A.~Cornia, A.~Barra, A.~Jansen, E.~Rentschler and G.~Timco, \emph{{Chem. Eur.
  J.}}, {2002}, \textbf{{8}}, {277--285}\relax
\mciteBstWouldAddEndPuncttrue
\mciteSetBstMidEndSepPunct{\mcitedefaultmidpunct}
{\mcitedefaultendpunct}{\mcitedefaultseppunct}\relax
\EndOfBibitem
\bibitem[Larsen \emph{et~al.}({2003})Larsen, Overgaard, Parsons, Rentschler,
  Smith, Timco, and Winpenny]{LOP:ACIE03}
F.~Larsen, J.~Overgaard, S.~Parsons, E.~Rentschler, A.~Smith, G.~Timco and
  R.~Winpenny, \emph{{Angew. Chem. Int. Edit.}}, {2003}, \textbf{{42}},
  {5978--5981}\relax
\mciteBstWouldAddEndPuncttrue
\mciteSetBstMidEndSepPunct{\mcitedefaultmidpunct}
{\mcitedefaultendpunct}{\mcitedefaultseppunct}\relax
\EndOfBibitem
\bibitem[Larsen \emph{et~al.}({2003})Larsen, McInnes, El~Mkami, Rajaraman,
  Rentschler, Smith, Smith, Boote, Jennings, Timco, and Winpenny]{LMM:03}
F.~Larsen, E.~McInnes, H.~El~Mkami, G.~Rajaraman, E.~Rentschler, A.~Smith,
  G.~Smith, V.~Boote, M.~Jennings, G.~Timco and R.~Winpenny, \emph{{Angew.
  Chem. Int. Edit.}}, {2003}, \textbf{{42}}, {101--105}\relax
\mciteBstWouldAddEndPuncttrue
\mciteSetBstMidEndSepPunct{\mcitedefaultmidpunct}
{\mcitedefaultendpunct}{\mcitedefaultseppunct}\relax
\EndOfBibitem
\bibitem[Cador \emph{et~al.}(2004)Cador, Gatteschi, Sessoli, Larsen, Overgaard,
  Barra, Teat, Timco, and Winpenny]{CGS:ACIE04}
O.~Cador, D.~Gatteschi, R.~Sessoli, F.~K. Larsen, J.~Overgaard, A.~L. Barra,
  S.~J. Teat, G.~A. Timco and R.~E.~P. Winpenny, \emph{Angew. Chem. Int.
  Edit.}, 2004, \textbf{43}, 5196--5200\relax
\mciteBstWouldAddEndPuncttrue
\mciteSetBstMidEndSepPunct{\mcitedefaultmidpunct}
{\mcitedefaultendpunct}{\mcitedefaultseppunct}\relax
\EndOfBibitem
\bibitem[Cador \emph{et~al.}(2005)Cador, Gatteschi, Sessoli, Barra, Timco, and
  Winpenny]{CGS:JMMM05}
O.~Cador, D.~Gatteschi, R.~Sessoli, A.-L. Barra, G.~A. Timco and R.~E.~P.
  Winpenny, \emph{J. Magn. Magn. Mater.}, 2005, \textbf{290-291}, 55--60\relax
\mciteBstWouldAddEndPuncttrue
\mciteSetBstMidEndSepPunct{\mcitedefaultmidpunct}
{\mcitedefaultendpunct}{\mcitedefaultseppunct}\relax
\EndOfBibitem
\bibitem[Yao \emph{et~al.}(2006)Yao, Wang, Ma, Waldmann, Du, Song, Li, Zheng,
  Decurtins, and Xin]{YWM:CC06}
H.~C. Yao, J.~J. Wang, Y.~S. Ma, O.~Waldmann, W.~X. Du, Y.~Song, Y.~Z. Li,
  L.~M. Zheng, S.~Decurtins and X.~Q. Xin, \emph{Chem. Commun.}, 2006,
  1745--1747\relax
\mciteBstWouldAddEndPuncttrue
\mciteSetBstMidEndSepPunct{\mcitedefaultmidpunct}
{\mcitedefaultendpunct}{\mcitedefaultseppunct}\relax
\EndOfBibitem
\bibitem[Furukawa \emph{et~al.}(2009)Furukawa, Kiuchi, ichi Kumagai, Ajiro,
  Narumi, Iwaki, Kindo, Bianchi, Carretta, Santini, Borsa, Timco, and
  Winpenny]{FKK:PRB09}
Y.~Furukawa, K.~Kiuchi, K.~ichi Kumagai, Y.~Ajiro, Y.~Narumi, M.~Iwaki,
  K.~Kindo, A.~Bianchi, S.~Carretta, P.~Santini, F.~Borsa, G.~A. Timco and
  R.~E.~P. Winpenny, \emph{Phys. Rev. B}, 2009, \textbf{79}, 134416\relax
\mciteBstWouldAddEndPuncttrue
\mciteSetBstMidEndSepPunct{\mcitedefaultmidpunct}
{\mcitedefaultendpunct}{\mcitedefaultseppunct}\relax
\EndOfBibitem
\bibitem[Hoshino \emph{et~al.}(2009)Hoshino, Nakano, Nojiri, Wernsdorfer, and
  Oshio]{HNN:JACS09}
N.~Hoshino, M.~Nakano, H.~Nojiri, W.~Wernsdorfer and H.~Oshio, \emph{J. Am.
  Chem. Soc.}, 2009, \textbf{131}, 15100--15101\relax
\mciteBstWouldAddEndPuncttrue
\mciteSetBstMidEndSepPunct{\mcitedefaultmidpunct}
{\mcitedefaultendpunct}{\mcitedefaultseppunct}\relax
\EndOfBibitem
\bibitem[B{\"a}rwinkel \emph{et~al.}(2000)B{\"a}rwinkel, Schmidt, and
  Schnack]{BSS:JMMM00:B}
K.~B{\"a}rwinkel, H.-J. Schmidt and J.~Schnack, \emph{J. Magn. Magn. Mater.},
  2000, \textbf{220}, 227\relax
\mciteBstWouldAddEndPuncttrue
\mciteSetBstMidEndSepPunct{\mcitedefaultmidpunct}
{\mcitedefaultendpunct}{\mcitedefaultseppunct}\relax
\EndOfBibitem
\bibitem[B\"arwinkel \emph{et~al.}(2003)B\"arwinkel, Hage, Schmidt, and
  Schnack]{BHS:PRB03}
K.~B\"arwinkel, P.~Hage, H.-J. Schmidt and J.~Schnack, \emph{Phys. Rev. B},
  2003, \textbf{68}, 054422\relax
\mciteBstWouldAddEndPuncttrue
\mciteSetBstMidEndSepPunct{\mcitedefaultmidpunct}
{\mcitedefaultendpunct}{\mcitedefaultseppunct}\relax
\EndOfBibitem
\bibitem[Masuda \emph{et~al.}(2004)Masuda, Zheludev, Bush, Markina, and
  Vasiliev]{MZB:PRL04}
T.~Masuda, A.~Zheludev, A.~Bush, M.~Markina and A.~Vasiliev, \emph{Phys. Rev.
  Lett.}, 2004, \textbf{92}, 177201\relax
\mciteBstWouldAddEndPuncttrue
\mciteSetBstMidEndSepPunct{\mcitedefaultmidpunct}
{\mcitedefaultendpunct}{\mcitedefaultseppunct}\relax
\EndOfBibitem
\bibitem[Drechsler \emph{et~al.}({2005})Drechsler, Malek, Richter, Moskvin,
  Gippius, and Rosner]{DMR:PRL05}
S.~Drechsler, J.~Malek, J.~Richter, A.~Moskvin, A.~Gippius and H.~Rosner,
  \emph{Phys. Rev. Lett.}, {2005}, \textbf{{94}}, 039705\relax
\mciteBstWouldAddEndPuncttrue
\mciteSetBstMidEndSepPunct{\mcitedefaultmidpunct}
{\mcitedefaultendpunct}{\mcitedefaultseppunct}\relax
\EndOfBibitem
\bibitem[Richter \emph{et~al.}({1995})Richter, Ivanov, Retzlaff, and
  Voigt]{RIR:JMMM95}
J.~Richter, N.~Ivanov, K.~Retzlaff and A.~Voigt, \emph{J. Magn. Magn. Mater.},
  {1995}, \textbf{{140}}, {1611--1612}\relax
\mciteBstWouldAddEndPuncttrue
\mciteSetBstMidEndSepPunct{\mcitedefaultmidpunct}
{\mcitedefaultendpunct}{\mcitedefaultseppunct}\relax
\EndOfBibitem
\bibitem[Richter \emph{et~al.}(1995)Richter, Ivanov, Voigt, and
  Retzlaff]{RIV:JLTP95}
J.~Richter, N.~Ivanov, A.~Voigt and K.~Retzlaff, \emph{J.~Low Temp. Phys.},
  1995, \textbf{99}, 363\relax
\mciteBstWouldAddEndPuncttrue
\mciteSetBstMidEndSepPunct{\mcitedefaultmidpunct}
{\mcitedefaultendpunct}{\mcitedefaultseppunct}\relax
\EndOfBibitem
\bibitem[Henley({1989})]{Hen:PRL89}
C.~L. Henley, \emph{Phys. Rev. Lett.}, {1989}, \textbf{{62}},
  {2056--2059}\relax
\mciteBstWouldAddEndPuncttrue
\mciteSetBstMidEndSepPunct{\mcitedefaultmidpunct}
{\mcitedefaultendpunct}{\mcitedefaultseppunct}\relax
\EndOfBibitem
\bibitem[Waldtmann \emph{et~al.}(1998)Waldtmann, Everts, Bernu, Lhuillier,
  Sindzingre, Lecheminant, and Pierre]{WEB:EPJB98}
C.~Waldtmann, H.~U. Everts, B.~Bernu, C.~Lhuillier, P.~Sindzingre,
  P.~Lecheminant and L.~Pierre, \emph{Eur. Phys. J. B}, 1998, \textbf{2},
  501--507\relax
\mciteBstWouldAddEndPuncttrue
\mciteSetBstMidEndSepPunct{\mcitedefaultmidpunct}
{\mcitedefaultendpunct}{\mcitedefaultseppunct}\relax
\EndOfBibitem
\bibitem[Berg \emph{et~al.}(2003)Berg, Altman, and Auerbach]{BAA:PRL03}
E.~Berg, E.~Altman and A.~Auerbach, \emph{Phys. Rev. Lett.}, 2003, \textbf{90},
  147204\relax
\mciteBstWouldAddEndPuncttrue
\mciteSetBstMidEndSepPunct{\mcitedefaultmidpunct}
{\mcitedefaultendpunct}{\mcitedefaultseppunct}\relax
\EndOfBibitem
\bibitem[Schmidt \emph{et~al.}(2005)Schmidt, Schnack, and Richter]{SSR:JMMM05}
R.~Schmidt, J.~Schnack and J.~Richter, \emph{J. Magn. Magn. Mater.}, 2005,
  \textbf{295}, 164--167\relax
\mciteBstWouldAddEndPuncttrue
\mciteSetBstMidEndSepPunct{\mcitedefaultmidpunct}
{\mcitedefaultendpunct}{\mcitedefaultseppunct}\relax
\EndOfBibitem
\bibitem[Schnalle and Schnack(2009)]{ScS:PRB09}
R.~Schnalle and J.~Schnack, \emph{Phys. Rev. B}, 2009, \textbf{79},
  104419\relax
\mciteBstWouldAddEndPuncttrue
\mciteSetBstMidEndSepPunct{\mcitedefaultmidpunct}
{\mcitedefaultendpunct}{\mcitedefaultseppunct}\relax
\EndOfBibitem
\bibitem[Schnack and Schnalle(2009)]{ScS:P09}
J.~Schnack and R.~Schnalle, \emph{Polyhedron}, 2009, \textbf{28},
  1620--1623\relax
\mciteBstWouldAddEndPuncttrue
\mciteSetBstMidEndSepPunct{\mcitedefaultmidpunct}
{\mcitedefaultendpunct}{\mcitedefaultseppunct}\relax
\EndOfBibitem
\bibitem[Schnalle(2009)]{Schnalle:Diss09}
R.~Schnalle, \emph{Ph.D. thesis}, Osnabr{\"uck} University, 2009\relax
\mciteBstWouldAddEndPuncttrue
\mciteSetBstMidEndSepPunct{\mcitedefaultmidpunct}
{\mcitedefaultendpunct}{\mcitedefaultseppunct}\relax
\EndOfBibitem
\bibitem[Rousochatzakis \emph{et~al.}(2008)Rousochatzakis, L\"{a}uchli, and
  Mila]{RLM:PRB08}
I.~Rousochatzakis, A.~M. L\"{a}uchli and F.~Mila, \emph{Phys. Rev. B}, 2008,
  \textbf{77}, 094420\relax
\mciteBstWouldAddEndPuncttrue
\mciteSetBstMidEndSepPunct{\mcitedefaultmidpunct}
{\mcitedefaultendpunct}{\mcitedefaultseppunct}\relax
\EndOfBibitem
\bibitem[Azaria \emph{et~al.}(1998)Azaria, Hooley, Lecheminant, Lhuillier, and
  Tsvelik]{AHL:PRL98}
P.~Azaria, C.~Hooley, P.~Lecheminant, C.~Lhuillier and A.~M. Tsvelik,
  \emph{Phys. Rev. Lett.}, 1998, \textbf{81}, 1694--1697\relax
\mciteBstWouldAddEndPuncttrue
\mciteSetBstMidEndSepPunct{\mcitedefaultmidpunct}
{\mcitedefaultendpunct}{\mcitedefaultseppunct}\relax
\EndOfBibitem
\bibitem[Atwood(2002)]{Atw:NM02}
J.~L. Atwood, \emph{Nat. Mater.}, 2002, \textbf{1}, 91--92\relax
\mciteBstWouldAddEndPuncttrue
\mciteSetBstMidEndSepPunct{\mcitedefaultmidpunct}
{\mcitedefaultendpunct}{\mcitedefaultseppunct}\relax
\EndOfBibitem
\bibitem[Zhitomirsky(2002)]{Zhi:PRL02}
M.~E. Zhitomirsky, \emph{Phys. Rev. Lett.}, 2002, \textbf{88}, 057204\relax
\mciteBstWouldAddEndPuncttrue
\mciteSetBstMidEndSepPunct{\mcitedefaultmidpunct}
{\mcitedefaultendpunct}{\mcitedefaultseppunct}\relax
\EndOfBibitem
\bibitem[Honecker \emph{et~al.}(2002)Honecker, Petrenko, and
  Zhitomirsky]{HPZ:PB02}
A.~Honecker, O.~A. Petrenko and M.~E. Zhitomirsky, \emph{Physica B}, 2002,
  \textbf{312}, 609--611\relax
\mciteBstWouldAddEndPuncttrue
\mciteSetBstMidEndSepPunct{\mcitedefaultmidpunct}
{\mcitedefaultendpunct}{\mcitedefaultseppunct}\relax
\EndOfBibitem
\bibitem[Cabra \emph{et~al.}(2005)Cabra, Grynberg, Holdsworth, Honecker, Pujol,
  Richter, Schmalfu\ss{}, and Schulenburg]{CGH:PRB05}
D.~C. Cabra, M.~D. Grynberg, P.~C.~W. Holdsworth, A.~Honecker, P.~Pujol,
  J.~Richter, D.~Schmalfu\ss{} and J.~Schulenburg, \emph{Phys. Rev. B}, 2005,
  \textbf{71}, 144420\relax
\mciteBstWouldAddEndPuncttrue
\mciteSetBstMidEndSepPunct{\mcitedefaultmidpunct}
{\mcitedefaultendpunct}{\mcitedefaultseppunct}\relax
\EndOfBibitem
\bibitem[Schr{\"o}der \emph{et~al.}(2005)Schr{\"o}der, Nojiri, Schnack, Hage,
  Luban, and K{\"o}gerler]{SNS:PRL05}
C.~Schr{\"o}der, H.~Nojiri, J.~Schnack, P.~Hage, M.~Luban and P.~K{\"o}gerler,
  \emph{Phys. Rev. Lett.}, 2005, \textbf{94}, 017205\relax
\mciteBstWouldAddEndPuncttrue
\mciteSetBstMidEndSepPunct{\mcitedefaultmidpunct}
{\mcitedefaultendpunct}{\mcitedefaultseppunct}\relax
\EndOfBibitem
\bibitem[Kawamura and Miyashita(1985)]{KaM:JPSJ85}
H.~Kawamura and S.~Miyashita, \emph{J. Phys. Soc. Jpn.}, 1985, \textbf{54},
  4530--4538\relax
\mciteBstWouldAddEndPuncttrue
\mciteSetBstMidEndSepPunct{\mcitedefaultmidpunct}
{\mcitedefaultendpunct}{\mcitedefaultseppunct}\relax
\EndOfBibitem
\bibitem[Penc \emph{et~al.}(2004)Penc, Shannon, and Shiba]{PSS:PRL04}
K.~Penc, N.~Shannon and H.~Shiba, \emph{Phys. Rev. Lett.}, 2004, \textbf{93},
  197203\relax
\mciteBstWouldAddEndPuncttrue
\mciteSetBstMidEndSepPunct{\mcitedefaultmidpunct}
{\mcitedefaultendpunct}{\mcitedefaultseppunct}\relax
\EndOfBibitem
\bibitem[Schnack \emph{et~al.}(2001)Schnack, Schmidt, Richter, and
  Schulenburg]{SSR:EPJB01}
J.~Schnack, H.-J. Schmidt, J.~Richter and J.~Schulenburg, \emph{Eur. Phys. J.
  B}, 2001, \textbf{24}, 475\relax
\mciteBstWouldAddEndPuncttrue
\mciteSetBstMidEndSepPunct{\mcitedefaultmidpunct}
{\mcitedefaultendpunct}{\mcitedefaultseppunct}\relax
\EndOfBibitem
\bibitem[Lhuillier and Misguich(2002)]{LhM:02}
C.~Lhuillier and G.~Misguich, in \emph{High Magnetic Fields}, ed. C.~Berthier,
  L.~Levy and G.~Martinez, Springer, Berlin, 2002, pp. 161--190\relax
\mciteBstWouldAddEndPuncttrue
\mciteSetBstMidEndSepPunct{\mcitedefaultmidpunct}
{\mcitedefaultendpunct}{\mcitedefaultseppunct}\relax
\EndOfBibitem
\bibitem[Schulenburg \emph{et~al.}(2002)Schulenburg, Honecker, Schnack,
  Richter, and Schmidt]{SHS:PRL02}
J.~Schulenburg, A.~Honecker, J.~Schnack, J.~Richter and H.-J. Schmidt,
  \emph{Phys. Rev. Lett.}, 2002, \textbf{88}, 167207\relax
\mciteBstWouldAddEndPuncttrue
\mciteSetBstMidEndSepPunct{\mcitedefaultmidpunct}
{\mcitedefaultendpunct}{\mcitedefaultseppunct}\relax
\EndOfBibitem
\bibitem[Richter \emph{et~al.}(2004)Richter, Schulenburg, Honecker, Schnack,
  and Schmidt]{RSH:JPCM04}
J.~Richter, J.~Schulenburg, A.~Honecker, J.~Schnack and H.-J. Schmidt, \emph{J.
  Phys.: Condens. Matter}, 2004, \textbf{16}, S779\relax
\mciteBstWouldAddEndPuncttrue
\mciteSetBstMidEndSepPunct{\mcitedefaultmidpunct}
{\mcitedefaultendpunct}{\mcitedefaultseppunct}\relax
\EndOfBibitem
\bibitem[Zhitomirsky and Tsunetsugu(2005)]{ZhT:IKYS05}
M.~E. Zhitomirsky and H.~Tsunetsugu, \emph{Prog. Theor. Phys. Suppl.}, 2005,
  \textbf{160}, 361--382\relax
\mciteBstWouldAddEndPuncttrue
\mciteSetBstMidEndSepPunct{\mcitedefaultmidpunct}
{\mcitedefaultendpunct}{\mcitedefaultseppunct}\relax
\EndOfBibitem
\bibitem[Szymczak \emph{et~al.}(2006)Szymczak, Baran, Diduszko, Fink-Finowicki,
  Gutowska, Szewcyk, and Szymczak]{SBD:PRB06}
R.~Szymczak, M.~Baran, R.~Diduszko, J.~Fink-Finowicki, M.~Gutowska, A.~Szewcyk
  and H.~Szymczak, \emph{Phys. Rev. B}, 2006, \textbf{73}, 094425\relax
\mciteBstWouldAddEndPuncttrue
\mciteSetBstMidEndSepPunct{\mcitedefaultmidpunct}
{\mcitedefaultendpunct}{\mcitedefaultseppunct}\relax
\EndOfBibitem
\bibitem[Rule \emph{et~al.}({2008})Rule, Wolter, Suellow, Tennant, Bruehl,
  Koehler, Wolf, Lang, and Schreuer]{RWS:PRL08}
K.~C. Rule, A.~U.~B. Wolter, S.~Suellow, D.~A. Tennant, A.~Bruehl, S.~Koehler,
  B.~Wolf, M.~Lang and J.~Schreuer, \emph{{Phys. Rev. Lett.}}, {2008},
  \textbf{{100}}, {117202}\relax
\mciteBstWouldAddEndPuncttrue
\mciteSetBstMidEndSepPunct{\mcitedefaultmidpunct}
{\mcitedefaultendpunct}{\mcitedefaultseppunct}\relax
\EndOfBibitem
\bibitem[Derzhko and Richter(2004)]{DeR:PRB04}
O.~Derzhko and J.~Richter, \emph{Phys. Rev. B}, 2004, \textbf{70}, 104415\relax
\mciteBstWouldAddEndPuncttrue
\mciteSetBstMidEndSepPunct{\mcitedefaultmidpunct}
{\mcitedefaultendpunct}{\mcitedefaultseppunct}\relax
\EndOfBibitem
\bibitem[Schmidt \emph{et~al.}(2006)Schmidt, Richter, and Moessner]{SRM:JPA06}
H.-J. Schmidt, J.~Richter and R.~Moessner, \emph{J. Phys. A: Math. Gen.}, 2006,
  \textbf{39}, 10673--10690\relax
\mciteBstWouldAddEndPuncttrue
\mciteSetBstMidEndSepPunct{\mcitedefaultmidpunct}
{\mcitedefaultendpunct}{\mcitedefaultseppunct}\relax
\EndOfBibitem
\bibitem[Zhitomirsky and Tsunetsugu(2007)]{ZhT:PRB07}
M.~E. Zhitomirsky and H.~Tsunetsugu, \emph{Phys. Rev. B}, 2007, \textbf{75},
  224416\relax
\mciteBstWouldAddEndPuncttrue
\mciteSetBstMidEndSepPunct{\mcitedefaultmidpunct}
{\mcitedefaultendpunct}{\mcitedefaultseppunct}\relax
\EndOfBibitem
\bibitem[Derzhko \emph{et~al.}(2007)Derzhko, Honecker, and Richter]{DHR:PRB07}
O.~Derzhko, A.~Honecker and J.~Richter, \emph{Phys. Rev. B}, 2007, \textbf{76},
  220402\relax
\mciteBstWouldAddEndPuncttrue
\mciteSetBstMidEndSepPunct{\mcitedefaultmidpunct}
{\mcitedefaultendpunct}{\mcitedefaultseppunct}\relax
\EndOfBibitem
\bibitem[Tasaki(1992)]{Tas:PRL92}
H.~Tasaki, \emph{Phys. Rev. Lett.}, 1992, \textbf{69}, 1608--1611\relax
\mciteBstWouldAddEndPuncttrue
\mciteSetBstMidEndSepPunct{\mcitedefaultmidpunct}
{\mcitedefaultendpunct}{\mcitedefaultseppunct}\relax
\EndOfBibitem
\bibitem[Mielke(1992)]{Mie:JPA92A}
A.~Mielke, \emph{J. Phys. A-Math. Gen.}, 1992, \textbf{25}, 6507--6515\relax
\mciteBstWouldAddEndPuncttrue
\mciteSetBstMidEndSepPunct{\mcitedefaultmidpunct}
{\mcitedefaultendpunct}{\mcitedefaultseppunct}\relax
\EndOfBibitem
\bibitem[Mielke(1992)]{Mie:JPA92B}
A.~Mielke, \emph{J. Phys. A-Math. Gen.}, 1992, \textbf{25}, 4335--4345\relax
\mciteBstWouldAddEndPuncttrue
\mciteSetBstMidEndSepPunct{\mcitedefaultmidpunct}
{\mcitedefaultendpunct}{\mcitedefaultseppunct}\relax
\EndOfBibitem
\bibitem[Mielke and Tasaki(1993)]{MiT:CMP93}
A.~Mielke and H.~Tasaki, \emph{Commun. Math. Phys.}, 1993, \textbf{158},
  341--371\relax
\mciteBstWouldAddEndPuncttrue
\mciteSetBstMidEndSepPunct{\mcitedefaultmidpunct}
{\mcitedefaultendpunct}{\mcitedefaultseppunct}\relax
\EndOfBibitem
\bibitem[Zhitomirsky and Honecker(2004)]{ZhH:JSM04}
M.~E. Zhitomirsky and A.~Honecker, \emph{J. Stat. Mech.: Theor. Exp.}, 2004,
  P07012\relax
\mciteBstWouldAddEndPuncttrue
\mciteSetBstMidEndSepPunct{\mcitedefaultmidpunct}
{\mcitedefaultendpunct}{\mcitedefaultseppunct}\relax
\EndOfBibitem
\bibitem[Schnack \emph{et~al.}(2007)Schnack, Schmidt, and Richter]{SSR:PRB07}
J.~Schnack, R.~Schmidt and J.~Richter, \emph{Phys. Rev. B}, 2007, \textbf{76},
  054413\relax
\mciteBstWouldAddEndPuncttrue
\mciteSetBstMidEndSepPunct{\mcitedefaultmidpunct}
{\mcitedefaultendpunct}{\mcitedefaultseppunct}\relax
\EndOfBibitem
\bibitem[Honecker and Zhitomirsky(2009)]{HoZ:JPCS09}
A.~Honecker and M.~E. Zhitomirsky, \emph{J. Phys.: Conf. Ser.}, 2009,
  \textbf{145}, 012082 (4pp)\relax
\mciteBstWouldAddEndPuncttrue
\mciteSetBstMidEndSepPunct{\mcitedefaultmidpunct}
{\mcitedefaultendpunct}{\mcitedefaultseppunct}\relax
\EndOfBibitem
\bibitem[Zhitomirsky(2003)]{Zhi:PRB03}
M.~E. Zhitomirsky, \emph{Phys. Rev. B}, 2003, \textbf{67}, 104421\relax
\mciteBstWouldAddEndPuncttrue
\mciteSetBstMidEndSepPunct{\mcitedefaultmidpunct}
{\mcitedefaultendpunct}{\mcitedefaultseppunct}\relax
\EndOfBibitem
\bibitem[McMichael \emph{et~al.}(1992)McMichael, Shull, Swartzendruber,
  Bennett, and Watson]{MSS:JMMM92}
R.~D. McMichael, R.~D. Shull, L.~J. Swartzendruber, L.~H. Bennett and R.~E.
  Watson, \emph{J. Magn. Magn. Mater.}, 1992, \textbf{111}, 29--33\relax
\mciteBstWouldAddEndPuncttrue
\mciteSetBstMidEndSepPunct{\mcitedefaultmidpunct}
{\mcitedefaultendpunct}{\mcitedefaultseppunct}\relax
\EndOfBibitem
\bibitem[Bennett \emph{et~al.}(1994)Bennett, McMichael, Tang, and
  Watson]{BMT:JAP94}
L.~H. Bennett, R.~D. McMichael, H.~C. Tang and R.~E. Watson, \emph{J. Appl.
  Phys.}, 1994, \textbf{75}, 5493--5495\relax
\mciteBstWouldAddEndPuncttrue
\mciteSetBstMidEndSepPunct{\mcitedefaultmidpunct}
{\mcitedefaultendpunct}{\mcitedefaultseppunct}\relax
\EndOfBibitem
\bibitem[Affronte \emph{et~al.}(2004)Affronte, Ghirri, Carretta, Amoretti,
  Piligkos, Timco, and Winpenny]{AGC:APL04}
M.~Affronte, A.~Ghirri, S.~Carretta, G.~Amoretti, S.~Piligkos, G.~A. Timco and
  R.~E.~P. Winpenny, \emph{Appl. Phys. Lett.}, 2004, \textbf{84},
  3468--3470\relax
\mciteBstWouldAddEndPuncttrue
\mciteSetBstMidEndSepPunct{\mcitedefaultmidpunct}
{\mcitedefaultendpunct}{\mcitedefaultseppunct}\relax
\EndOfBibitem
\bibitem[Evangelisti \emph{et~al.}(2005)Evangelisti, Candini, Ghirri, Affronte,
  Brechin, and McInnes]{ECG:APL05}
M.~Evangelisti, A.~Candini, A.~Ghirri, M.~Affronte, E.~K. Brechin and E.~J.
  McInnes, \emph{Appl. Phys. Lett.}, 2005, \textbf{87}, 072504\relax
\mciteBstWouldAddEndPuncttrue
\mciteSetBstMidEndSepPunct{\mcitedefaultmidpunct}
{\mcitedefaultendpunct}{\mcitedefaultseppunct}\relax
\EndOfBibitem
\bibitem[Evangelisti \emph{et~al.}(2009)Evangelisti, Candini, Affronte, Pasca,
  de~Jongh, Scott, and Brechin]{evangelisti:104414}
M.~Evangelisti, A.~Candini, M.~Affronte, E.~Pasca, L.~J. de~Jongh, R.~T.~W.
  Scott and E.~K. Brechin, \emph{Phys. Rev. B}, 2009, \textbf{79}, 104414\relax
\mciteBstWouldAddEndPuncttrue
\mciteSetBstMidEndSepPunct{\mcitedefaultmidpunct}
{\mcitedefaultendpunct}{\mcitedefaultseppunct}\relax
\EndOfBibitem
\bibitem[Coffey and Trugman(1992)]{CoT:PRL92}
D.~Coffey and S.~A. Trugman, \emph{Phys. Rev. Lett.}, 1992, \textbf{69},
  176--179\relax
\mciteBstWouldAddEndPuncttrue
\mciteSetBstMidEndSepPunct{\mcitedefaultmidpunct}
{\mcitedefaultendpunct}{\mcitedefaultseppunct}\relax
\EndOfBibitem
\bibitem[Schr{\"o}der \emph{et~al.}(2005)Schr{\"o}der, Schmidt, Schnack, and
  Luban]{SSS:PRL05}
C.~Schr{\"o}der, H.-J. Schmidt, J.~Schnack and M.~Luban, \emph{Phys. Rev.
  Lett.}, 2005, \textbf{94}, 207203\relax
\mciteBstWouldAddEndPuncttrue
\mciteSetBstMidEndSepPunct{\mcitedefaultmidpunct}
{\mcitedefaultendpunct}{\mcitedefaultseppunct}\relax
\EndOfBibitem
\bibitem[Konstantinidis(2007)]{Kon:PRB07}
N.~P. Konstantinidis, \emph{Phys. Rev. B}, 2007, \textbf{76}, 104434\relax
\mciteBstWouldAddEndPuncttrue
\mciteSetBstMidEndSepPunct{\mcitedefaultmidpunct}
{\mcitedefaultendpunct}{\mcitedefaultseppunct}\relax
\EndOfBibitem
\bibitem[Micotti \emph{et~al.}(2006)Micotti, Furukawa, Kumagai, Carretta,
  Lascialfari, Borsa, Timco, and Winpenny]{MFK:PRL06}
E.~Micotti, Y.~Furukawa, K.~Kumagai, S.~Carretta, A.~Lascialfari, F.~Borsa,
  G.~A. Timco and R.~E.~P. Winpenny, \emph{Phys. Rev. Lett.}, 2006,
  \textbf{97}, 267204\relax
\mciteBstWouldAddEndPuncttrue
\mciteSetBstMidEndSepPunct{\mcitedefaultmidpunct}
{\mcitedefaultendpunct}{\mcitedefaultseppunct}\relax
\EndOfBibitem
\bibitem[Corradini \emph{et~al.}(2008)Corradini, Moro, Biagi, del Pennino,
  Renzi, Carretta, Santini, Affronte, Cezar, Timco, and Winpenny]{CMB:PRB08}
V.~Corradini, F.~Moro, R.~Biagi, U.~del Pennino, V.~D. Renzi, S.~Carretta,
  P.~Santini, M.~Affronte, J.~C. Cezar, G.~Timco and R.~E.~P. Winpenny,
  \emph{Phys. Rev. B}, 2008, \textbf{77}, 014402\relax
\mciteBstWouldAddEndPuncttrue
\mciteSetBstMidEndSepPunct{\mcitedefaultmidpunct}
{\mcitedefaultendpunct}{\mcitedefaultseppunct}\relax
\EndOfBibitem
\bibitem[Saalfrank \emph{et~al.}(2006)Saalfrank, Scheurer, Bernt, Heinemann,
  Postnikov, Sch{\"u}nemann, Trautwein, Alam, Rupp, and M{\"u}ller]{SSB:DT06}
R.~W. Saalfrank, A.~Scheurer, I.~Bernt, F.~W. Heinemann, A.~V. Postnikov,
  V.~Sch{\"u}nemann, A.~X. Trautwein, M.~S. Alam, H.~Rupp and P.~M{\"u}ller,
  \emph{Dalton Trans.}, 2006,  2865--2874\relax
\mciteBstWouldAddEndPuncttrue
\mciteSetBstMidEndSepPunct{\mcitedefaultmidpunct}
{\mcitedefaultendpunct}{\mcitedefaultseppunct}\relax
\EndOfBibitem
\bibitem[Khanra \emph{et~al.}(2007)Khanra, Biswas, Golze, B{\"u}chner, Kataev,
  Weyhermuller, and Chaudhuri]{KBG:DT07}
S.~Khanra, B.~Biswas, C.~Golze, B.~B{\"u}chner, V.~Kataev, T.~Weyhermuller and
  P.~Chaudhuri, \emph{Dalton Trans.}, 2007,  481--487\relax
\mciteBstWouldAddEndPuncttrue
\mciteSetBstMidEndSepPunct{\mcitedefaultmidpunct}
{\mcitedefaultendpunct}{\mcitedefaultseppunct}\relax
\EndOfBibitem
\bibitem[Prinz \emph{et~al.}()Prinz, Kuepper, Taubitz, Raekers, Biswas,
  Weyherm{\"u}ller, Uhlarz, Wosnitza, Schnack, Postnikov, Schr{\"o}der, George,
  Neumann, , and Chaudhuri]{PKT:09}
M.~Prinz, K.~Kuepper, C.~Taubitz, M.~Raekers, B.~Biswas, T.~Weyherm{\"u}ller,
  M.~Uhlarz, J.~Wosnitza, J.~Schnack, A.~V. Postnikov, C.~Schr{\"o}der, S.~J.
  George, M.~Neumann,  and P.~Chaudhuri, \emph{Inorg. Chem.}\relax
\mciteBstWouldAddEndPunctfalse
\mciteSetBstMidEndSepPunct{\mcitedefaultmidpunct}
{}{\mcitedefaultseppunct}\relax
\EndOfBibitem
\bibitem[Tidmarsh \emph{et~al.}(2009)Tidmarsh, Batchelor, Scales, Laye, Sorace,
  Caneschi, Schnack, and McInnes]{TBS:DT09}
I.~S. Tidmarsh, L.~J. Batchelor, E.~Scales, R.~H. Laye, L.~Sorace, A.~Caneschi,
  J.~Schnack and E.~J.~L. McInnes, \emph{Dalton Trans.}, 2009,
  9402--9409\relax
\mciteBstWouldAddEndPuncttrue
\mciteSetBstMidEndSepPunct{\mcitedefaultmidpunct}
{\mcitedefaultendpunct}{\mcitedefaultseppunct}\relax
\EndOfBibitem
\bibitem[Schnack(2000)]{Schnack:PRB00}
J.~Schnack, \emph{Phys. Rev. B}, 2000, \textbf{62}, 14855\relax
\mciteBstWouldAddEndPuncttrue
\mciteSetBstMidEndSepPunct{\mcitedefaultmidpunct}
{\mcitedefaultendpunct}{\mcitedefaultseppunct}\relax
\EndOfBibitem
\bibitem[Schollw\"ock \emph{et~al.}(2004)Schollw\"ock, Richter, Farnell, and
  Bishop]{SRF:LNP04}
\emph{Quantum Magnetism}, ed. U.~Schollw\"ock, J.~Richter, D.~Farnell and
  R.~Bishop, Springer, Berlin, Heidelberg, 2004, vol. 645\relax
\mciteBstWouldAddEndPuncttrue
\mciteSetBstMidEndSepPunct{\mcitedefaultmidpunct}
{\mcitedefaultendpunct}{\mcitedefaultseppunct}\relax
\EndOfBibitem
\bibitem[Diep(2005)]{Diep05}
\emph{Frustrated spin systems}, ed. H.~Diep, World Scientific, Singapore,
  2005\relax
\mciteBstWouldAddEndPuncttrue
\mciteSetBstMidEndSepPunct{\mcitedefaultmidpunct}
{\mcitedefaultendpunct}{\mcitedefaultseppunct}\relax
\EndOfBibitem
\bibitem[Lacroix \emph{et~al.}(2010)Lacroix, Mila, and Mendels]{LMM:10}
\emph{Introduction to Frustrated Magnetism}, ed. C.~Lacroix, F.~Mila and
  P.~Mendels, Springer, Berlin \& Heidelberg, 2010\relax
\mciteBstWouldAddEndPuncttrue
\mciteSetBstMidEndSepPunct{\mcitedefaultmidpunct}
{\mcitedefaultendpunct}{\mcitedefaultseppunct}\relax
\EndOfBibitem
\bibitem[Misguich \emph{et~al.}(1998)Misguich, Bernu, Lhuillier, and
  Waldtmann]{MBL:PRL98}
G.~Misguich, B.~Bernu, C.~Lhuillier and C.~Waldtmann, \emph{Phys. Rev. Lett.},
  1998, \textbf{81}, 1098--1101\relax
\mciteBstWouldAddEndPuncttrue
\mciteSetBstMidEndSepPunct{\mcitedefaultmidpunct}
{\mcitedefaultendpunct}{\mcitedefaultseppunct}\relax
\EndOfBibitem
\bibitem[Manaka \emph{et~al.}(2009)Manaka, Hirai, Hachigo, Mitsunaga, Ito, and
  Terada]{JPSJ.78.093701}
H.~Manaka, Y.~Hirai, Y.~Hachigo, M.~Mitsunaga, M.~Ito and N.~Terada, \emph{J.
  Phys. Soc. Jpn.}, 2009, \textbf{78}, 093701\relax
\mciteBstWouldAddEndPuncttrue
\mciteSetBstMidEndSepPunct{\mcitedefaultmidpunct}
{\mcitedefaultendpunct}{\mcitedefaultseppunct}\relax
\EndOfBibitem
\bibitem[F\aa{}k \emph{et~al.}(2008)F\aa{}k, Coomer, Harrison, Visser, and
  Zhitomirsky]{0295-5075-81-1-17006}
B.~F\aa{}k, F.~C. Coomer, A.~Harrison, D.~Visser and M.~E. Zhitomirsky,
  \emph{Europhys. Lett.}, 2008, \textbf{81}, 17006\relax
\mciteBstWouldAddEndPuncttrue
\mciteSetBstMidEndSepPunct{\mcitedefaultmidpunct}
{\mcitedefaultendpunct}{\mcitedefaultseppunct}\relax
\EndOfBibitem
\bibitem[Bramwell \emph{et~al.}(2009)Bramwell, Giblin, Calder, Aldus,
  Prabhakaran, and Fennell]{BGC:Nature09}
S.~T. Bramwell, S.~R. Giblin, S.~Calder, R.~Aldus, D.~Prabhakaran and
  T.~Fennell, \emph{Nature}, 2009, \textbf{461}, 956--959\relax
\mciteBstWouldAddEndPuncttrue
\mciteSetBstMidEndSepPunct{\mcitedefaultmidpunct}
{\mcitedefaultendpunct}{\mcitedefaultseppunct}\relax
\EndOfBibitem
\end{mcitethebibliography}

\providecommand*{\mcitethebibliography}{\thebibliography}
\csname @ifundefined\endcsname{endmcitethebibliography}
{\let\endmcitethebibliography\endthebibliography}{}

\end{document}